\documentclass[12pt]{article}
\usepackage{psfig}

\makeatletter
\def\vereq#1#2{\lower3pt\vbox{\baselineskip1.5pt \lineskip1.5pt
\ialign{$\m@th#1\hfill##\hfil$\crcr#2\crcr\sim\crcr}}}
\makeatother

\begin{document}

\begin{titlepage}
\noindent
\begin{flushright}
CERN-TH/2000-241 \\
hep-ph/0008085 \\
\end{flushright}

\vspace{0.5cm}
\begin{center}
  \begin{Large}
  \begin{bf}
    Lepton-Flavour Violation in Supersymmetric Models with
   Trilinear R-parity Violation
   \\
  \end{bf}
 \end{Large}
\end{center}
  \vspace{0.5cm}
 
    \begin{center}
      Andr\'e de Gouv\^ea, Smaragda Lola, and Kazuhiro Tobe\\
      \vspace{0.3cm}
\begin{it}
CERN -- Theory Division\\
CH-1211 Gen\`eve 23, Switzerland
\end{it}

  \end{center}

\vspace{0.5cm}

\begin{abstract}
Supersymmetry with R-parity violation (RPV) provides an interesting 
framework for naturally accommodating small neutrino masses. 
Within this framework, we discuss the 
lepton-flavour violating (LFV) processes $\mu \rightarrow e \gamma$, $\mu 
\rightarrow eee$, and $\mu \rightarrow e$ conversion in nuclei.
We make a detailed study of the observables related to LFV in different RPV models,
and compare them to the expectations of R-conserving supersymmetry with heavy 
right-handed neutrinos. We show that the predictions are vastly different
and uniquely characterise each model, thus providing a powerful framework for 
experimentally distinguishing between different theories of LFV.
Besides the obvious possibility of amplified tree-level 
generation of $\mu \rightarrow eee$ and $\mu \rightarrow e$ conversion in nuclei,
we find that even in the case where these processes arise at the one-loop level, 
their rates are comparable to that of $\mu\rightarrow e\gamma$, in clear
contrast to the predictions of R-conserving models. We conclude that in order to 
distinguish between the different models, such a combined study of {\em all} the 
LFV processes is necessary, and that measuring P-odd asymmetries in polarised $\mu 
\rightarrow eee$ can play a decisive role. We also comment on the intriguing 
possibility of RPV models yielding a large T-odd asymmetry in the 
decay of polarised $\mu\rightarrow eee$.
\end{abstract}

\end{titlepage}

\newpage
\setcounter{equation}{0}
\section{Introduction}

Recently, neutrino oscillation experiments~\cite{atm,solar,LSND} have provided
very strong evidence for non-zero, yet tiny, neutrino masses. 
In order to accommodate such small masses, it is widely believed that new physics 
beyond the Standard Model (SM) is required. One of the simplest and most elegant 
mechanisms for generating a small neutrino mass is to introduce extra standard
model singlets to the SM Lagrangian, and allow them to acquire a very large Majorana
mass (this is the well known seesaw mechanism \cite{seesaw}). There are many 
important phenomenological consequences of neutrino masses. One of them is that
individual lepton-flavour numbers are not conserved, which implies that SM forbidden
processes such as $\mu\rightarrow e\gamma$ may occur. However, given the
size of the neutrino masses, the rates for charged lepton flavour violating (LFV) 
phenomena are extremely small in the SM plus massive neutrinos \cite{neutrinoLFVns}.  

There are other hints for physics beyond the SM, including the gauge hierarchy 
problem. Low-energy supersymmetry (SUSY) is one of the preferred candidates for 
beyond the SM physics which solves the hierarchy problem. SUSY models can easily 
accomodate the seesaw mechanism, and SUSY even helps in the sense that it stabilises 
the (very heavy) Majorana mass of the right-handed neutrino. Furthermore, in such a 
framework, LFV processes in the charged lepton sector such as $\mu \rightarrow e 
\gamma$, $\mu \rightarrow eee$, and $\mu \rightarrow e$ conversion in nuclei are
potentially amplified, as has been previously discussed 
\cite{masiero, tobe, hisano, Okada_kuno},
and the rates for such processes can be within the reach of future experiments. 
The reason for this is that while in the SM plus massive neutrinos the amplitudes for
LFV violation are proportional to the neutrino masses ({\it i.e.,}\/ suppressed by 
the very large right-handed neutrino masses, in the case of the seesaw mechanism), 
in SUSY models these processes are only suppressed by inverse powers of the 
supersymmetry breaking scale, which is at most $O(1)$~TeV.

Another SM extension which naturally accommodates non-zero neutrino masses 
is SUSY with R-parity violation (RPV). R-parity is usually imposed as a global
symmetry of the minimal supersymmetric version of the SM (MSSM) in order to prevent 
an unacceptably large rate for proton decay. 
However, this proves to be somewhat of an overkill, since R-parity conservation 
implies both baryon number and lepton number conservation, while to stop proton
decay only one or the other needs to be exactly conserved.  
In light of the evidence for neutrino masses, which can potentially be
Majorana particles and therefore violate lepton number, one may, instead, 
take advantage of RPV operators to generate small neutrino masses. 

In this paper, we consider SUSY models with RPV but with baryon parity (in order
to satisfy the current experimental upper limits on the proton 
lifetime~\cite{ROSS}).\footnote{In cosmology, large RPV Yukawa couplings may erase a
pre-existing baryon asymmetry~\cite{Fukugita}. Here we do not
consider such constraints since they are model-dependent and can be evaded in 
several baryogenesis scenarios
\cite{EWBaryon}.} These models naturally generate small Majorana neutrino masses,
if the RPV couplings are small \cite{neutrino_Rp,pila,bhat}.\footnote{A mechanism 
which explains why RPV couplings are small is required. This can be achieved, for 
example, by imposing flavour symmetries which relate the lepton and baryon number 
violating Yukawa couplings to those that generate fermion masses \cite{MODELS}.} 
In such RPV models, ``large'' LFV in the charged lepton sector is also
 generically expected. Indeed, as has been pointed out in the literature
\cite{Choudhury,Chaichian,Kim,Huitu,Faessler,Choi},
the most stringent limits on certain products of RPV couplings
come from the present experimental bounds on charged LFV processes.  
Therefore it is important to understand some of the general features of LFV in
models with RPV.

It is interesting to consider how searches for LFV at low energy experiments 
compare to those at colliders. For instance, the similtaneous presence of 
R-violating operators that couple both to $e-q$ and
to $\mu -q$ ($\tau -q$) pairs, would lead to $\mu$+jet ($\tau$+jet) final states at 
HERA \cite{HERA}. 
It turns out that for $e \leftrightarrow \tau$ transitions, the high energy 
experimental probes provide the strongest bounds, while for $e \leftrightarrow 
\mu$ transitions stopped muon experiments provide, by far, the most stringent bounds.
Finally, the strongest bound for $\tau\leftrightarrow\mu$ transitions comes from 
$\tau \rightarrow \mu \gamma$ searches at CLEO \cite{CLEO} ($Br(\tau \rightarrow 
\mu \gamma) < 1.1 \times  10^{-6}$) which is less restrictive. In the near future,
the experimental sensitivity to some rare muons processes is going to improve
by two to three orders of magnitude, while a similar improvement is not expected 
for other LFV processes. For this reason, we will focus on processes with stopped 
muons, which not only provide the stringest quantitative bounds on LFV today, 
but which will be significantly probed in the near future.

In this paper, we discuss the LFV processes $\mu^+ \rightarrow e^+ \gamma$,
$\mu^+ \rightarrow e^+e^-e^+$, and $\mu^- \rightarrow e^-$ conversion in the case
of models with trilinear RPV. In Sec.~2, we briefly introduce the SUSY models with 
trilinear RPV which will be considered here. 
In Sec.~3, we present the formalism for computing branching ratios
and asymmetries of the relevant LFV processes. In Sec.~4, we consider LFV processes 
in some representative cases, including those in which the branching ratio for 
$\mu^+\rightarrow e^+\gamma$ is much smaller than the
branching ratio for $\mu^+\rightarrow e^+e^-e^+$ and/or the rate for $\mu^-\rightarrow
e^-$ conversion in nuclei, which can be generated at the 
tree-level. Even if all LFV processes occur at the one-loop level, the rates for
all three processes considered here are comparable. These features are completely 
different from the predictions of other neutrino mass generating SUSY frameworks, 
such as the MSSM with right-handed neutrinos \cite{tobe}. In the latter, 
the branching ratio for $\mu^+\rightarrow e^+\gamma$ is much larger than that for 
$\mu^+\rightarrow e^+e^-e^+$ and the rate for $\mu^-\rightarrow e^-$ conversion in 
nuclei, even though all processes are also generated at the one-loop 
level. We also show that P-odd asymmetries in 
the $\mu^+ \rightarrow e^+e^-e^+$ process (which require polarised muons in order to
be measured) are very useful in order to distinguish different models. 
Sec.~5 contains our conclusions. In Appendix~A we provide 
explicit expressions for the LFV vertices in the case of models with RPV, while in 
Appendix~B we discuss the current bounds on certain pairs of RPV couplings from
LFV processes and comment on neutrino masses. 

\setcounter{equation}{0}
\section{SUSY models with trilinear R-parity violation}

Here, we briefly introduce the SUSY models with RPV which will be discussed
 in the upcoming sections. If R-parity conservation is not postulated, in addition to 
ordinary Yukawa interactions, the following terms are allowed in the MSSM 
superpotential:
\begin{eqnarray}
W_{RPV}&=&\frac{\lambda_{ijk}}{2}L_i L_j \bar{E}_k + \lambda '_{ijk}L_i Q_j 
\bar{D}_k +
\lambda ''_{ijk} \bar{U}_i \bar{D}_j \bar{D}_k + \mu'_{i} L_i H_u,
\label{wrpv}
\end{eqnarray}
where $L_i$, $\bar{E}_i$, $Q_i$, $\bar{U}_i$, $\bar{D}_i$, and $H_u$ 
denote the left-handed doublet lepton,
right-handed lepton, left-handed doublet quark, right-handed up-type quark, 
right-handed down-type quark, and 
``up-type'' Higgs superfields, respectively. The indices $i,j$ and $k$ range from 
1 to 3 for different quark/lepton flavours. Throughout this paper, in order to 
forbid rapid proton decay, we impose baryon parity~\cite{ROSS}, so all $\lambda ''$ 
couplings are zero. We also make the simplifying assumption that all $\mu'$ also 
vanish.\footnote{Even if $\mu'_i$ were non-zero, their contributions to LFV processes
would be, in general, negligible because of neutrino mass constraints \cite{pila,
Choi}.} In light of these assumptions, the superpotential above yields the following 
Lagrangian:
\begin{eqnarray}
{\cal L}&=& \lambda_{ijk} \left(
\bar{\nu}^c_{Li} e_{Lj} \tilde{e}_{Rk}^*
+\bar{e}_{Rk} \nu_{Li} \tilde{e}_{Lj}
+\bar{e}_{Rk} e_{Lj} \tilde{\nu}_{Li}
\right)
\nonumber \\
&+&\lambda '_{ijk}V_{KM}^{j\alpha} \left(
\bar{\nu}^c_{Li} d_{L\alpha} \tilde{d}^*_{Rk}
+ \bar{d}_{Rk} \nu_{Li} \tilde{d}_{L\alpha}
+ \bar{d}_{Rk} d_{L\alpha} \tilde{\nu}_{Li} \right)
\nonumber \\
&-&\lambda '_{ijk} \left(
\bar{u}^c_{j} e_{Li} \tilde{d}^*_{Rk} +
\bar{d}_{Rk} e_{Li} \tilde{u}_{Lj}
+ \bar{d}_{Rk} u_{Lj} \tilde{e}_{Li} \right)
+{\rm h.c.},
\label{RPVmodel}
\end{eqnarray}
where $f$ ($f=\nu,e,d$, and $u$) denotes fermions and $\tilde{f}$
sfermions, and the index $(R,L)$ indicates the field's chirality.
We assume that the RPV Yukawa couplings above ($\lambda_{ijk}$ and $\lambda '_{ijk}$) 
are the only source of LFV. In what follows, Eq.~(\ref{RPVmodel}) is what is 
referred to by ``RPV model.''  

\setcounter{equation}{0}
\section{Branching ratios and asymmetries for the LFV processes}

In this section, we present complete expressions for the branching ratios 
for the LFV processes $\mu^+ \rightarrow e^+ \gamma$, $\mu^+ \rightarrow 
e^+e^-e^+$, and $\mu^- \rightarrow e^-$ conversion in nuclei, for the P-odd asymmetry
in $\mu^+\rightarrow e^+\gamma$, and for the P-odd and
T-odd asymmetries in $\mu^+ \rightarrow e^+e^-e^+$.

\subsection{$\mu^+ \rightarrow e^+ \gamma$}
The process $\mu^+ \rightarrow e^+ \gamma^{(*)}$ is generated by 
photon penguin diagrams (see the penguin diagrams in Figs.~\ref{diagram1},
\ref{diagram2}, and \ref{diagram3}). The amplitude for this process can be written as 
follows:
\begin{eqnarray}
T &=&e \epsilon^{\alpha *} \bar{v}_{\mu}(p)
\left[ (A_1^L P_L + A_1^R P_R) 
\gamma^\beta (g_{\alpha \beta} q^2 - q_\alpha q_\beta)
\right.
\nonumber 
\\ &&\left. \hspace{4cm}+ m_\mu i \sigma_{\alpha \beta} q^\beta
(A_2^L P_L + A_2^R P_R)
\right] v_e (p-q),
\label{photon_penguin}
\end{eqnarray}
where $v_{\mu(e)}$ and $\epsilon$ are the antimuon (positron) and photon
wave functions, and $p$ and $q$ are the antimuon and photon momenta, respectively.
$P_L$ and $P_R$ are chirality projection operators: $P_L=(1-\gamma_5)/2$,
and $P_R=(1+\gamma_5)/2$, while $\sigma_{\alpha \beta}=(i/2)
[\gamma_\alpha,~\gamma_\beta]$. The effective couplings $A_{1}^{L,R}$ come from 
off-shell photon diagrams ($q^2 \neq 0$), which only contribute to $\mu^+ 
\rightarrow e^+e^-e^+$ and $\mu\rightarrow e$ conversion in nuclei. 
On the other hand, the couplings $A_{2}^{L,R}$ arise from the on-shell photon
diagrams ($q^2=0$), which induce $\mu^+ \rightarrow e^+ \gamma$ as well as 
$\mu^+ \rightarrow e^+e^-e^+$ and $\mu^- \rightarrow e^-$ conversion in nuclei.
Explicit expressions for $A_{1,2}^{L,R}$ in models with RPV are presented in 
Appendix~A.

In the $\mu^+ \rightarrow e^+ \gamma$ decay, it has been argued \cite{kuno}
that a nonzero muon polarisation is useful not only to 
suppress background processes, but also to distinguish between
$\mu^+ \rightarrow e^+_L \gamma$ and $\mu^+ \rightarrow e^+_R \gamma$.
The differential branching ratio for $\mu^+ \rightarrow e^+ \gamma$ is given by
\begin{eqnarray}
\frac{d{\rm Br}(\mu^+ \rightarrow e^+ \gamma)}{d \cos\theta}
&=&\frac{{\rm Br}(\mu^+ \rightarrow e^+ \gamma)}{2}
\left\{ 1+  A_P P \cos \theta \right\},
\end{eqnarray}
where $P$ is the muon polarisation and $\theta$ is the angle between the 
positron momentum and the polarisation direction. Here,
the branching ratio Br$(\mu^+ \rightarrow e^+ \gamma)$ and the P-odd asymmetry 
$A_P$ are
\begin{eqnarray}
{\rm{Br}}(\mu^+ \rightarrow e^+ \gamma) &=& \frac{48 \pi^3 \alpha}{G_F^2}
\left(
|A_2^L |^2 + |A_2^R |^2
\right),
\\
A_P &=& \frac{|A_2^L |^2 - |A_2^R |^2}
{|A_2^L |^2 + |A_2^R |^2},
\label{APgamma}
\end{eqnarray}
where $G_F$ is the Fermi constant, and $\alpha$ is the fine-structure constant.

\subsection{Polarised $\mu^+ \rightarrow e^+ e^+ e^-$}

In the RPV models, some of the $LL\bar{E}$ couplings ($\lambda_{ijk}$)
generate $\mu^+ \rightarrow e^+e^-e^+$ at tree-level (Fig.~\ref{diagram1}), while
the photon penguin vertices $A_{1,2}^{L,R}$ also contribute.\footnote{There is also 
a Z-penguin contribution. However, its contribution is suppressed by
$m^2_f/m^2_Z$ where $m_f$ is the typical fermion mass in the process. 
Therefore we simply neglect it. In order to be consistent, we will not study processes
where top-quarks are involved.} The amplitude for $\mu^+ \rightarrow e^+e^-e^+$ is
\begin{eqnarray}
T &=& B^L \bar{v}_{\mu}(p) P_L \gamma_{\mu}v_e(p_2) 
\bar{u}_e(p_3) P_R\gamma^{\mu} v_e(p_1) \nonumber \\
&+&B^R \bar{v}_{\mu}(p) P_R\gamma_{\mu} v_e(p_2)
\bar{u}_e(p_3) P_L\gamma^{\mu} v_e(p_1)
\nonumber \\
&&\hspace{-1cm} +4 \pi \alpha
\bar{v}_{\mu} (p) \left\{ (A_1^L P_L+A_1^R P_R) \gamma_{\mu} 
+m_\mu i \frac{\sigma_{\mu \nu} q^\nu}{q^2} (A_2^R P_R+A_2^L P_L) \right\}
v_e(p_2)
\nonumber \\
&&\times \bar{u}_e(p_3) \gamma^\mu v_e(p_1) - 
(p_1 \leftrightarrow p_2),
\label{B_vertex}
\end{eqnarray}
where the explicit expressions for the tree-level vertices $B^{R(L)}$ in models 
with RPV are given in Appendix~A.

When the muon is polarised, two P-odd and one T-odd asymmetry
can be defined \cite{TWZ,Okada}. Using the notation introduced by Okada {\it et 
al.}\/~\cite{Okada}, the $z$-axis is taken to be the direction of the electron 
momentum and the $(z\times x)$--plane is taken to be the decay plane. 
The positron with the largest energy is denoted as positron 1 and the other as 
positron 2. The $x$-coordinate is defined as $(p_1)_x \ge 0$ where $\vec{p}_1$ is the 
momentum of positron 1. It is in this coordinate system that the direction of the  
muon polarisation $\vec{P}$, used below, is defined. (For details, see \cite{Okada}.)
Finally, the P-odd and T-odd asymmetries are defined as follows:
\begin{eqnarray}
A_{P_1} &=& \frac{N(P_{\it z} >0)-N(P_{\it z} <0)}
{N(P_{\it z} >0)+N(P_{\it z} <0)},
\nonumber 
\\ &=& \frac{3}{2 {\rm Br}(\delta)}
 \left\{ 0.61 (C_1-C_2)-0.12 (C_3-C_4) +5.6 (C_5-C_6) \right.\nonumber \\
     &&\left. \hspace{3cm}-4.7 (C_7-C_8)+2.5 (C_9-C_{10}) \right\},
\label{p-odd1}
\\
A_{P_2} &=& \frac{N(P_{\it x} >0)-N(P_{\it x} <0)}
{N(P_{\it x} >0)+N(P_{\it x} <0)},
\nonumber 
\\ &=& \frac{3}{2 {\rm Br}(\delta)}
 \left\{0.1 (C_3-C_4) +10 (C_5-C_6) \right. \nonumber \\
&&\left.\hspace{3cm}+2.0 (C_7-C_8) -1.6 (C_9-C_{10}) \right\},
\label{p-odd2}
\\
A_{T} &=& \frac{N(P_{\it y} >0)-N(P_{\it y} <0)}
{N(P_{\it y} >0)+N(P_{\it y} <0)},
\nonumber 
\\ &=& \frac{3}{2 {\rm Br}(\delta)}
\left\{2.0 C_{11}-1.6 C_{12} \right\},
\label{t-odd}
\end{eqnarray}
where the muons are assumed to be $100\%$ polarised, and  $N(P_i >(<)0)$ 
denotes the number of events with a positive (negative) $P_i$ component for the muon 
polarisation. Here an energy cutoff for positron 1 is introduced 
($E_1< (m_\mu/2)(1-\delta)$) and henceforth we will consider 
$\delta=0.02$, following Okada {\it et al.}\/ \cite{Okada}. This choice is made
in order to optimise the T-odd asymmetry. 
Of course, one can obtain more 
information concerning the $C_i$ coefficients, including the CP-odd terms $C_{11}$
and $C_{12}$, (see definition of $C_i$ in what follows) 
by analysing the Dalitz plot of the $\mu^+\rightarrow e^+e^-e^+$ decay.
$C_i$ ($i=1-12$) are functions of the effective couplings $A_{1,2}^{L,R}$ and 
$B^{L,R}$:
\begin{eqnarray}
C_1 &=& \frac{2 \pi^2 \alpha^2}{G_F^2} |A_1^R|^2,~~
C_2=\frac{2 \pi^2 \alpha^2}{G_F^2} |A_1^L|^2, 
\label{c12}
\\
C_3 &=& \frac{1}{8 G_F^2} |B^R +4\pi \alpha A_1^R|^2,~~
C_4 = \frac{1}{8 G_F^2} |B^L +4\pi \alpha A_1^L|^2, 
\label{c34}
\\
C_5 &=& \frac{\pi^2 \alpha^2}{2 G_F^2} |A_2^R|^2,~~
C_6  = \frac{\pi^2 \alpha^2}{2 G_F^2} |A_2^L|^2,\\
C_7 &=& -\frac{\pi^2 \alpha^2}{G_F^2} {\rm Re} (A_2^R A_1^{L*}),~~
C_8 = -\frac{\pi^2 \alpha^2}{G_F^2} {\rm Re} (A_2^L A_1^{R*}),\\
C_9 &=& -\frac{\pi \alpha}{4 G_F^2} {\rm Re} \{ A_2^R
(B^{L*}+4\pi \alpha A_1^{L*})\}, \\
C_{10} &=& -\frac{\pi \alpha}{4 G_F^2} {\rm Re} \{ A_2^L
(B^{R*}+4\pi \alpha A_1^{R*})\}, \\
C_{11} &=& \frac{\pi \alpha}{8 G_F^2}
{\rm Im}\left\{ 8\pi \alpha (A_2^{R} A_1^{L*} + A_2^{L} A_1^{R*})
\right\},
\\
C_{12} &=& \frac{\pi\alpha}{4G_F^2}
{\rm Im} \left\{ A_2^{R}(B^{L*}+4 \pi \alpha A_1^{*L})
+A_2^{L}(B^{R*} + 4 \pi \alpha A_1^{R*}) \right\}.
\end{eqnarray}
The branching ratio for $\delta=0.02$ is
\begin{eqnarray}
{\rm Br}(\delta=0.02) &=& 1.8 (C_1+C_2)+0.96 (C_3+C_4)
+88 (C_5+C_6)  \nonumber \\
&& +14 (C_7+C_8) +8 (C_9+C_{10}).
\label{opt_br}
\end{eqnarray}
The branching ratio for $\mu^+ \rightarrow e^+e^-e^+$ for $\delta=0$ is given by
\begin{eqnarray}
{\rm{Br}}(\mu^+ \rightarrow e^+e^-e^+) &=& 2(C_1+C_2) +C_3+C_4+
32 \left\{\log\frac{m_\mu^2}{m_e^2}-\frac{11}{4}\right\} (C_5+C_6)
\nonumber \\
&&+16(C_7+C_8) + 8 (C_9+C_{10})
\\
&=& \frac{1}{8 G_F^2}\left[
|B^L|^2+|B^R|^2 +48 \pi^2 \alpha^2
\left\{ |A_1^L|^2+|A_1^R|^2 
\right. \right.
\nonumber\\
&&\left.\hspace{-2cm}+\frac{8}{3}
\left(\log \frac{m^2_\mu}{m^2_e}-\frac{11}{4}\right)
(|A_2^R|^2+|A_2^L|^2)
-4{\rm{Re}}(A_1^L A_2^{R*}+A_1^R A_2^{L*})\right\}
\nonumber \\
&&\left. \hspace{-1cm}
+8\pi \alpha {\rm{Re}}\left\{A_1^L B^{L*}+A_1^R B^{R*}
-2(A_2^R B^{L*}+A_2^L B^{R*})\right\}
\right].
\label{br(mueee)}
\end{eqnarray}

\subsection{$\mu^- \rightarrow e^-$ conversion in nuclei}
Similarly to $\mu^+\rightarrow e^+e^-e^+$, not only photon penguin diagrams but also 
tree-level diagram induced by some of the $LQ\bar{D}$ Yukawa couplings 
($\lambda '_{ijk}$) can generate $\mu^- \rightarrow e^-$ conversion in nuclei. The 
amplitude is given by 
\begin{eqnarray}
T &=& D^u 
\bar{u}_e \gamma_\mu P_L u_\mu~ \bar{u}_{u} \gamma_\mu P_L u_{u} 
+ D^d 
\bar{u}_e \gamma_\mu P_L u_\mu~\bar{u}_{d} \gamma_\mu P_R u_{d} 
\nonumber \\
&-&4 \pi \alpha
\bar{u}_e \left\{ \gamma_\mu (A_1^{L*} P_L+A_1^{R*} P_R) 
+m_\mu i \frac{\sigma_{\mu \nu} q^\nu}{q^2} (A_2^{R*} P_R+A_2^{L*} P_L) \right\}
u_\mu
\nonumber \\
&&\times \sum_{q=u,d}Q_q \bar{u}_q \gamma_\mu u_q,
\label{D_vertex}
\end{eqnarray}
where the complete expressions for the tree-level contributions $D^{u,d}$ in the 
case of RPV models are presented in Appendix~A. The $\mu^- \rightarrow e^-$ 
conversion rate is
\begin{eqnarray}
{\rm R}(\mu^- \rightarrow e^-)
&=&\frac{\alpha^3 Z_{eff}^4 |F(q)|^2 m_\mu^5}
{16 \pi^2 Z ~\Gamma(\mu~ {\rm capture})} 
\left\{ 64 \pi^2 \alpha^2 |A_1^R-A_2^L|^2 
\right. \\
&+&\left.\left| (2Z+N) D^u+(Z+2N) D^d -8 \pi \alpha Z 
(A_1^{L*}-A_2^{R*})\right|^2 \right\}, \nonumber
\end{eqnarray}
where $\Gamma(\mu~ {\rm capture})$ is the muon capture rate in the 
nucleus of interest \cite{capture}, $Z$ and $N$ are the proton and neutron numbers, 
respectively, $F(q)$ is the nuclear form factor as a function of the momentum transfer
and $Z_{eff}$ is the nuclear effective charge \cite{mu_e_conv}.
In some of the most commonly used nuclei, $^{48}_{22}{\rm Ti}$ and $^{27}_{13}{\rm Al}$,
these nuclear parameters are given by \cite{mu_e_conv}
\begin{eqnarray}
&&\Gamma(\mu~ {\rm capture})
=2.590\times10^6~s^{-1}=1.7\times10^{-18}~{\rm GeV},
\nonumber \\
&& Z=22,~Z_{eff}=17.61,~|F(q^2=-m_\mu^2)|=0.535~~ {\rm for}~^{48}_{22}{\rm Ti},
\end{eqnarray}
and
\begin{eqnarray}
&&\Gamma(\mu~ {\rm capture})
=0.7054\times10^6~s^{-1}=4.6\times10^{-19}~{\rm GeV},
\nonumber \\
&& Z=13,~Z_{eff}=11.62,~|F(q^2=-m_\mu^2)|=0.64~~{\rm for}~^{27}_{13}{\rm Al}.
\end{eqnarray}

\setcounter{equation}{0}
\section{LFV in representative cases}

The most severe constraints on some particular products of trilinear RPV
couplings come from the present experimental upper limits on the branching ratios 
of the LFV processes discussed in the previous section (see Appendix~B)
\cite{Choudhury,Chaichian,Kim,Huitu,Faessler,Choi}.  
Therefore, searches for LFV in muon processes are particularly sensitive
to models with RPV. Generically, it is very hard to make definite predictions
for the branching ratios of the LFV processes since the number of new Yukawa couplings
($\lambda_{ijk},\lambda '_{ijk}$) is too large. Here we consider, instead, different 
cases where only a small number of RPV couplings is significant for LFV. This is done
not only to simplify the problem at hand, but also to identify features of LFV 
which are not only different from those in the ``traditional'' models, such
as the MSSM with heavy right-handed neutrinos discussed in \cite{masiero,tobe,hisano},
but which can also be used to characterise the different cases themselves. This 
dominance of specific RPV couplings is also a consequence of certain flavour models 
\cite{MODELS}. 

\subsection{$\mu^+ \rightarrow e^+e^-e^+$ induced at tree-level}

\begin{figure}[t]
\centerline{
~~~~~~~~~\makebox(100,120)[t]{\psfig{figure=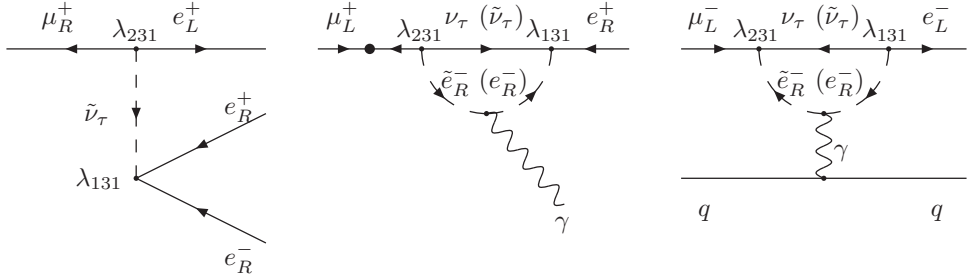,width=1.5\textwidth}}
}
\vspace*{0.4 cm}
\caption{\sl Lowest order Feynman diagrams for lepton flavour violating processes induced by
$\lambda_{131} \lambda_{231}$ couplings (see Eq.~(\ref{wrpv})). }
\label{diagram1}
\end{figure}

First, we consider a model in which only the Yukawa couplings 
$\lambda_{131}$ and $\lambda_{231}$ are non-zero. In this case, $\mu^+ 
\rightarrow e^+e^-e^+$
is generated at the tree-level, while the other LFV processes ($\mu^+ \rightarrow e^+ 
\gamma$ and $\mu^- \rightarrow e^-$ conversion in nuclei)
are induced via photon penguin diagrams at the one-loop level, as shown in 
Fig.~\ref{diagram1}.

The effective vertices are given by
\begin{eqnarray}
B^L &=& -\frac{\lambda_{131} \lambda_{231}}{2m^2_{\tilde{\nu}_\tau}},
\\
A_2^R &=& -\frac{\lambda_{131} \lambda_{231}}{96 \pi^2 m^2_{\tilde{\nu}_\tau}}
\left(1-\frac{m^2_{\tilde{\nu}_\tau}}{2m^2_{\tilde{e}_{R}}}\right),
\\
A_1^L &=& \frac{\lambda_{131} \lambda_{231}}{96 \pi^2 m^2_{\tilde{\nu}_\tau}}
\left(-\frac{8}{3}-2\log{\frac{m_e^2}{m^2_{\tilde{\nu}_\tau}}}
+\frac{m^2_{\tilde{\nu}_\tau}}{3 m^2_{\tilde{e}_R}}-2 \delta(m_e^2/q^2)
\right).
\end{eqnarray}
Here we assume, without loss of generality, that the RPV couplings are real.
The function $\delta$ is presented in Appendix~A.
In $\mu^- \rightarrow e^-$ conversion, we assume the
momentum of the virtual photon to be $q^2=-m^2_\mu$ in order to compute
$\delta(m_e^2/q^2)$ in $A_1^L$, while in the case of
$\mu^+ \rightarrow e^+e^-e^+$, we simply set
$q^2=0$ $(\delta=0)$, since the tree-level contribution $B^L$ is
much larger than the contribution from $\delta(m_e^2/q^2)$. The
ratios of branching ratios, 
${\rm Br}(\mu^+ \rightarrow e^+ \gamma)/{\rm Br}(\mu^+ \rightarrow e^+e^-e^+)$
and ${\rm R}(\mu^- \rightarrow e^- {\rm ~in ~nuclei})/{\rm Br}
(\mu^+ \rightarrow e^+e^-e^+)$
do not depend on the R-parity violating couplings $\lambda_{131} \lambda_{231}$, 
and only depend on the SUSY mass spectrum through $m^2_{\tilde{\nu}_\tau}$ and
$m^2_{\tilde{e}_R}$:
\begin{eqnarray}
\frac{{\rm Br}(\mu^+ \rightarrow e^+ \gamma)}{{\rm Br}(\mu^+ \rightarrow e^+e^-e^+)}
&=& \frac{4\times 10^{-4} \left(1-\frac{m^2_{\tilde{\nu}_\tau}}
{2 m^2_{\tilde{e}_R}} \right)^2}{\beta} =1\times 10^{-4},
\label{model1_ratio1}
\\
\frac{{\rm R}(\mu^- \rightarrow e^- {\rm ~in ~Ti~(Al)})}{{\rm Br}
(\mu^+ \rightarrow e^+e^-e^+)}
&=&\frac{2~(1) \times 10^{-5}}{\beta}
\left(\frac{5}{6} +\frac{m^2_{\tilde{\nu}_\tau}}{12 m^2_{\tilde{e}_R}}
+\log\frac{m^2_e}{m^2_{\tilde{\nu}_\tau}} +\delta \right)^2,
\nonumber \\
&=& 2~(1)\times 10^{-3},
\label{model1_ratio2}
\end{eqnarray}
the second of the equal signs being valid for $m_{\tilde{\nu}_\tau}=m_{\tilde{e}_R}=
100~{\rm GeV}$. Here $\beta=1+$(one-loop contr.)/(tree-level contr.) in the 
$\mu^+ \rightarrow e^+e^-e^+$ process, which is close to unity (for example, 
$\beta=0.98$ for $m_{\tilde{\nu}_\tau}=m_{\tilde{e}_R} =100~{\rm GeV}$).
Since the $\mu^+ \rightarrow e^+e^-e^+$ process is generated at tree level,
its branching ratio is much larger than that of the other LFV
processes, as expected. If such a scenario were realized in nature, 
the $\mu^+ \rightarrow e^+e^-e^+$ process would dominate over all
the other channels, {\it i.e.,}\/ it is very likely that if nature realizes this
particular scenario, $\mu^+ \rightarrow e^+e^-e^+$ is within experimental reach
while $\mu^+\rightarrow e^+\gamma$ 
is
orders of magnitude below any foreseeable future experiment.   

Another interesting feature of Eqs.(\ref{model1_ratio1},\ref{model1_ratio2})
is that, because of an ultraviolet log-enhancement of the off-shell photon penguin 
contribution ($A_1^L$)~\cite{Huitu,Sanda}, the $\mu^- \rightarrow e^-$ conversion rates 
are significantly larger than the branching ratio of $\mu^+ \rightarrow e^+\gamma$.

It is important to emphasise that the ratios of branching ratios of the different 
processes are very different from those in the different neutrino-mass models.
For example, in the MSSM with heavy right-handed neutrinos
(and R-parity conservation)~\cite{tobe}, the following relations are
approximately satisfied, because the on-shell photon penguin
contribution $A_2^R$ tends to dominate over all others:
\begin{eqnarray}
\frac{{\rm Br}(\mu^+ \rightarrow e^+ \gamma)}{{\rm Br}(\mu^+ \rightarrow e^+e^-e^+)}
 &\simeq& \frac{3\pi}{\alpha \left(\log\frac{m_\mu^2}{m_e^2}-\frac{11}{4}
\right)} =1.6 \times 10^2,
\label{ratioMSSM1}
\\
\frac{{\rm R}(\mu^- \rightarrow e^-~{{\rm in~Ti}})}
{{\rm Br}(\mu^+ \rightarrow e^+e^-e^+)} &\simeq& 
\frac{\alpha^3 Z_{eff}^4 Z |F(q)|^2 m_\mu^5 G_F^2}
{4\pi^2 \left(\log\frac{m_\mu^2}{m_e^2}-\frac{11}{4}
\right)\Gamma(\mu~{\rm capture})} = 0.92
\label{ratioMSSM2}
\end{eqnarray}

Another interesting feature of the case at hand is that in the $\mu^+ \rightarrow 
e^+e^-e^+$ process we obtain the following P-odd asymmetries 
(Eqs.~(\ref{p-odd1}-\ref{p-odd2})):
\begin{eqnarray}
A_{P_1} &\simeq& \frac{3(0.12 C_4)}{2 (0.96 C_4)}
= 19 \%,
\\
A_{P_2} &\simeq& \frac{-3 (0.1 C_4)}{2 (0.96 C_4)}
= -15 \%,
\\
\frac{A_{P_1}}{A_{P_2}} &\simeq& -1.3,
\end{eqnarray}
since the tree-level contribution $B^L$ ($C_4$) is dominant. The key feature here
is that the two different P-odd asymmetries
have opposite sign; $A_{P_1}/A_{P_2} \simeq -1.3$. 
More generally, this feature is present whenever the effective vertices 
$B^{L,R}$ are dominant. In Table~\ref{table1} we list results of other similar examples.

The situation is clearly different from the MSSM with heavy right-handed neutrinos,
where the on-shell photon contributions $A_2^R$ ($C_5$) are dominant 
(this case is also listed in Table~\ref{table1}, in order to facilitate comparisons):
\begin{eqnarray}
A_{P_1} &\simeq& \frac{3 (5.6 C_5)}{2 (87 C_5)}=10\%,
~~
A_{P_2} \simeq \frac{3 (10 C_5)}{2 (87 C_5)}=17\%,
\nonumber \\
\frac{A_{P_1}}{A_{P_2}} &\simeq& 0.6.
\end{eqnarray}
\begin{table}
\caption{\sl The ratios of branching ratios ${\rm Br}(\mu^+ \rightarrow e^+ \gamma)/
{\rm Br}(\mu^+ \rightarrow e^+e^-e^+)$ and 
${\rm R}(\mu^- \rightarrow e^-~{\rm in}~{\rm Ti})/{\rm Br}(\mu^+ 
\rightarrow e^+e^-e^+)$, P-odd asymmetries $A_P$ for $\mu^+ \rightarrow e^+ 
\gamma$, $A_{P_1}$ and $A_{P_2}$ for $\mu^+ \rightarrow e^+e^-e^+$ are shown
when the listed pair of Yukawa couplings is dominant. Case (1), (2), (3) refers to
the representative classes of models discussed in Secs.~4.1, 4.2, and 4,3, 
respectively. 
Here, we assume $m_{\tilde{\nu},\tilde{l}_R}=100$~GeV and no mixing in the 
charged slepton mass matrix, and $m_{\tilde{q}}=300$~GeV.
We also show a typical result obtained for the MSSM with heavy right-handed neutrinos
and R-parity conservation \cite{tobe}.}
\begin{center}
\begin{tabular}{|c|c|c|c|c|c|c|} \hline
& $\frac{{\rm Br}(\mu \rightarrow e \gamma)}{{\rm Br}(\mu \rightarrow 3e)}$ 
& $\frac{{\rm R}(\mu \rightarrow e~{\rm in}~{\rm Ti})}{{\rm Br}
(\mu \rightarrow 3e)}$ & $A_P$ & $A_{P_1}$ & $A_{P_2}$ & $A_{P_1}/A_{P_2}$ 
\\
\hline
Case (1) & & & & & & \\
$\lambda_{131} \lambda_{231}$ & $1\times 10^{-4}$ & $2\times 10^{-3}$ & $-100\%$ & $+19\%$
&$-15\%$ & $-1.3$\\
$\lambda_{121} \lambda_{122}$ & $8\times 10^{-4}$ & $7\times 10^{-3}$ & $+100\%$ & $-19\%$
&$+15\%$ & $-1.3$ \\
$\lambda_{131} \lambda_{132}$ & $8\times 10^{-4}$ & $5\times 10^{-3}$ & $+100\%$ & $-19\%$
&$+15\%$ & $-1.3$\\
\hline
Case (2) & & & & & & \\
$\lambda_{132} \lambda_{232}$ & $1.2$ & $18$ & $-100\%$ & $-25\%$ & $-5\%$ & $5.6$ \\
$\lambda_{133} \lambda_{233}$ & $3.7$ & $18$ & $-100\%$ & $-25\%$ & $-4\%$ & $6.2$ \\
$\lambda_{231} \lambda_{232}$ & $3.6$ & $18$ & $+100\%$ & $+25\%$ & $+4\%$ & $6.2$ \\
$\lambda '_{122} \lambda '_{222}$ & $1.4$ & $18$ & $-100\%$ & $-25\%$ & $-4\%$ & $5.7$\\
$\lambda '_{123} \lambda '_{223}$ & $2.2$ & $18$ & $-100\%$ & $-25\%$ & $-4\%$ & $5.9$\\
\hline
Case (3) & & & & & & \\
$\lambda '_{111} \lambda '_{211}$ & $0.4$ & $3\times10^2$ & $-100\%$ & $-26\%$ & $-5\%$ 
& $5.4$\\
$\lambda '_{112} \lambda '_{212}$ & $0.5$ & $8\times 10^{4}$ & $-100\%$ & $-26\%$ & $-5\%$ 
& $5.4$\\
$\lambda '_{113} \lambda '_{213}$ & $0.7$ & $1\times10^{5} $ & $-100\%$ & $-26\%$ & $-5\%$
& $5.5$\\
$\lambda '_{121} \lambda '_{221}$ & $1.1$ & $2\times10^{5}$ & $-100\%$ & $-26\%$ & $-5\%$
& $5.6$\\
\hline \hline
MSSM with $\nu_R$ & $1.6 \times 10^2$ & $0.92$ & $-100\%$ & $10\%$ & $17\%$ 
&$0.6$\\
\hline 
\end{tabular}
\end{center}
\label{table1}
\end{table}
Therefore, a measurement of the (sign of the) ratio of P-odd asymmetries in 
$\mu^+\rightarrow e^+e^-e^+$ can clearly separate these two models 
($B_{L,R}\gg A_i^{L,R}$ versus $A_{2}^{L,R}\gg B_{L,R}$). 

Another useful observable which may be measured in the case one has access to
polarised muon decays is $A_P$ (Eq.~(\ref{APgamma})). In RPV models, $A_P$ can
have different values (see Table~\ref{table1}), while in other SUSY extensions of
the SM, either $\mu^+\rightarrow e^+_L\gamma$ or $\mu^+\rightarrow e^+_R\gamma$ is 
forbidden. Some examples include R-parity conserving SUSY with right-handed handed
neutrinos (see Table~\ref{table1}), $SU(5)$ and $SO(10)$ SUSY grand unified theories,
and other MSSM extensions \cite{GUTs, Okada_kuno}. 

\subsection{All processes induced at one-loop level}
\begin{figure}[t]
\centerline{\hspace*{-1.2 cm}
\makebox(100,120)[t]{\psfig{figure=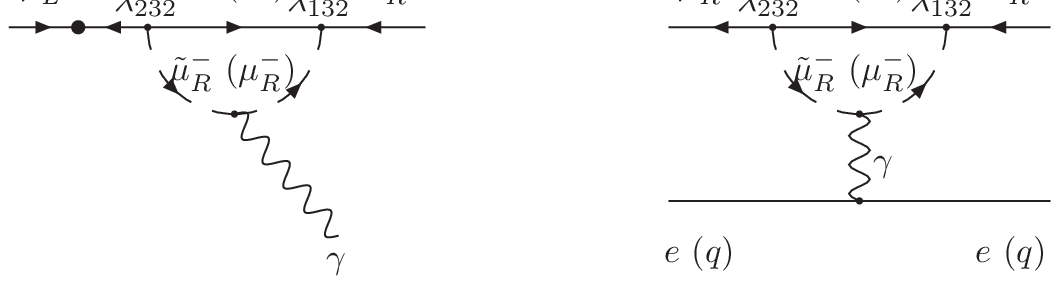,width=1.5\textwidth}}}
\vspace*{0.4cm}
\caption{\sl Lowest order Feynman diagrams for lepton flavour violating processes 
induced by $\lambda_{132} \lambda_{232}$ couplings (see Eq.~(\ref{wrpv})). }
\label{diagram2}
\end{figure}

Here we consider a different representative case,
in which all of $\mu^+ \rightarrow e^+ \gamma$,
$\mu^+ \rightarrow e^+e^-e^+$, and $\mu^- \rightarrow e^-$ conversion in nuclei are
induced at the one-loop level (at the lowest order in RPV couplings) through the 
photon penguin diagram (Fig.~\ref{diagram2}). Suppose, as an example, that only the 
couplings $\lambda_{132}$ and $\lambda_{232}$ are nonzero (again we assume that both 
of them are real, without loss of generality). 
The effective vertices for the LFV processes are 
\begin{eqnarray}
A_2^R &=& -\frac{\lambda_{132} \lambda_{232}}{96 \pi^2 m^2_{\tilde{\nu}_\tau}}
\left(1-\frac{m^2_{\tilde{\nu}_\tau}}{2 m^2_{\tilde{\mu}_{R}}}
\right),
\\
A_1^L &=&\frac{\lambda_{132} \lambda_{232}}{96 \pi^2 m^2_{\tilde{\nu}_\tau}}
\left(-\frac{8}{3} -2\log{\frac{m^2_{\mu}}{m^2_{\tilde{\nu}_\tau}}
+\frac{m^2_{\tilde{\nu}_\tau}}{3m^2_{\tilde{\mu}_{R}}}}
-\delta(m^2_\mu/q^2)
\right).
\label{case2}
\end{eqnarray}
Again, we set $q^2=-m^2_\mu$ for $\mu^- \rightarrow e^-$ conversion, and
$q^2=0$ for $\mu^+ \rightarrow e^+e^-e^+$.\footnote{Since the log-term is 
much larger than the $\delta$ term for $0<q^2<m^2_\mu$ in Eq.~(\ref{case2}), 
the result does not depend significantly on the choice of $q$ in the
$\mu^+ \rightarrow e^+e^-e^+$ process.}
The ratios of branching ratios ${\rm Br}(\mu^+ \rightarrow e^+ \gamma)/{\rm Br}
(\mu^+ \rightarrow e^+e^-e^+)$ and ${\rm R}(\mu^- \rightarrow e^-~{\rm in~nuclei})/
{\rm Br}(\mu^+ \rightarrow e^+e^-e^+)$ are independent on the choice of 
$\lambda_{132} \lambda_{232}$:
\begin{eqnarray}
\frac{{\rm Br}(\mu^+ \rightarrow e^+ \gamma)}{{\rm Br}(\mu^+ \rightarrow e^+e^-e^+)}
  &=& 3.2 \times 10^3
\frac{\left(1-\frac{m^2_{\tilde{\nu}_\tau}}{2m^2_{\tilde{\mu}_{R}}}\right)^2}
{\left(-\frac{8}{3}-2\log\frac{m^2_{\mu}}{m^2_{\tilde{\nu}_\tau}}
+\frac{m^2_{\tilde{\nu}_\tau}}{3m^2_{\tilde{\mu}_{R}}}\right)^2 \gamma},
\nonumber 
\\
 &=& 1.2,
\label{ratio1_case2}
\\
\frac{{\rm R}(\mu^- \rightarrow e^-~~{\rm in~Ti~(Al)})}
{{\rm Br}(\mu^+ \rightarrow e^+e^-e^+)}
&=& 19.5~(11.5) 
\frac{\left(\frac{5}{3}+\frac{m^2_{\tilde{\nu}_\tau}}{6m^2_{\tilde{\mu}_{R}}}
+2 \log\frac{m^2_{\mu}}{m^2_{\tilde{\nu}_\tau}}+\delta \right)^2}
{\left(-\frac{8}{3}-2\log\frac{m^2_{\mu}}{m^2_{\tilde{\nu}_\tau}}
+\frac{m^2_{\tilde{\nu}_\tau}}{3m^2_{\tilde{\mu}_{R}}}\right)^2 \gamma},
\nonumber \\
&=& 18~(11),
\label{ratio2_case2}
\end{eqnarray}
where the second of the equal signs holds for 
$m_{\tilde{\nu}_\tau}=m_{\tilde{\mu}_{R}}=100~{\rm GeV}$.
Here $\gamma$ is a function of the SUSY mass spectrum, but it is of order unity.
\begin{eqnarray}
\gamma=1+\frac{ \frac{8}{3} |A_2^R|^2 \left(\log \frac{m_\mu^2}{m_e^2}-
\frac{11}{4} \right)-4{\rm Re}(A_1^L A_2^R)}{|A_1^L|^2}.
\end{eqnarray}
As an example, $\gamma=1.09$ for $m_{\tilde{\nu}_\tau}
=m_{\tilde{\mu}_{R}}=100~{\rm GeV}$. 

\begin{figure}
\vspace*{-2.0 cm}
\centerline{
\makebox(100,400)[t]{\psfig{figure=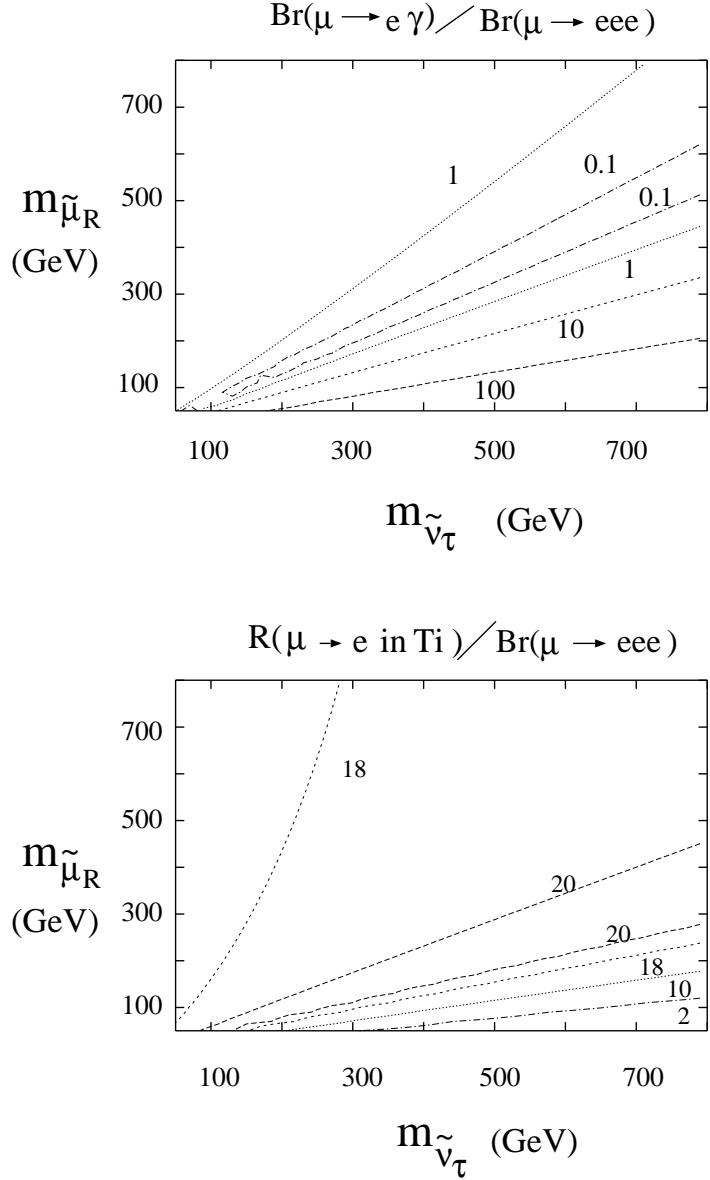,width=0.9\textwidth}}}
\vspace*{3.2 cm}
\caption{\sl Contours of constant Br$(\mu^+ \rightarrow e^+ \gamma)/$Br$(\mu^+ 
\rightarrow e^+e^-e^+)$
(top), and R$(\mu^- \rightarrow e^- ~{\rm in~Ti})/$Br$(\mu^+ \rightarrow e^+e^-e^+)$ 
(bottom) in the ($m_{\tilde{\mu}_R}\times m_{\tilde{\nu}_{\tau}}$)~plane, assuming
that only the product of $LL\bar{E}$ couplings $\lambda_{132} \lambda_{232}$ is 
non-zero (see Eq.~(\ref{wrpv})).}
\label{ratio_fig}
\end{figure}

Because of the ultraviolet log-enhancement of the off-shell photon penguin diagram 
($A_1^L$) in Eq.(\ref{case2}), the event rates for the $\mu^+ \rightarrow e^+e^-e^+$ 
and $\mu^- \rightarrow e^-$ conversion in nuclei can be as large as the branching 
ratio for the $\mu^+ \rightarrow e^+ \gamma$ process, even though they are higher
order processes in QED.\footnote{In the case of $\mu^+\rightarrow e^+e^-e^+$, there is
also an infrared log-enhancement to the branching ratio, as can be seen in 
Eq.~(\ref{br(mueee)}).} 
Fig.~\ref{ratio_fig} depicts the dependence on the slepton masses
of these ratios of branching ratios.
In the case of $\mu^+ \rightarrow e^+ \gamma$, a cancellation between the
two different diagrams (sneutrino and smuon loops) can occur, 
such that its branching ratio can be much smaller than that of the other processes. 
On the other hand, the numerical value of
the ratio R$(\mu^- \rightarrow e^- ~{\rm in~nuclei})/$Br$(\mu^+ 
\rightarrow e^+e^-e^+)$ is stable in a large region of the parameter space. 
All the LFV processes are equally relevant in this model. Again, we stress 
that these ratios of the branching ratios are very different in more ``traditional'' 
cases, such as in the MSSM with heavy right-handed neutrinos 
(see Eqs.(\ref{ratioMSSM1},\ref{ratioMSSM2})).

Since the off-shell photon diagram $A_1^L$ is dominant in the $\mu^+ \rightarrow 
e^+e^-e^+$ process, $C_2~(=C_4)$ in Eqs.(\ref{c12},\ref{c34}) is much larger than 
the other $C_i~(i \neq 2,4)$. In this case, the P-odd asymmetries behave as follows:
\begin{eqnarray}
A_{P_1} &\simeq& \frac{3 (-0.61 C_2 +0.12 C_4)}{2
(1.8 C_2 +0.96 C_4)}
=-26 \%
\\
A_{P_2} &\simeq& \frac{3(-0.1C_4)}{2
(1.8 C_2 + 0.96 C_4)}
=-5\%
\\
\frac{A_{P_2}}{A_{P_1}} &\simeq& 0.19.
\label{ratio3_case2}
\end{eqnarray}
These relations (Eqs.(\ref{ratio1_case2},\ref{ratio2_case2},\ref{ratio3_case2})) 
are a typical feature of models in which the off-shell photon diagram is the dominant 
contribution to the $\mu^+ \rightarrow e^-e^+e^-$ process.
The results of other similar examples are also listed in Table~\ref{table1}.

\subsection{$\mu^- \rightarrow e^-$ conversion in nuclei induced at tree-level}

\begin{figure}[t]
\centerline{
~~~~~~~~~\makebox(100,120)[t]{\psfig{figure=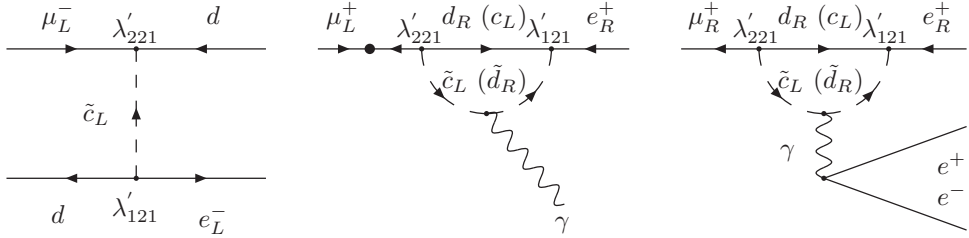,width=1.5\textwidth}}}
\vspace*{0.4cm}
\caption{\sl Lowest order Feynman diagrams of lepton flavour violating processes induced by
$f^{'}_{121} f^{'}_{221}$ couplings (see Eq.~(\ref{wrpv})). }
\label{diagram3}
\end{figure}

Here, we consider the possibility that $\mu^- \rightarrow e^-$ conversion in 
nuclei is induced at tree-level. This can arise through some of the $LQ\bar{D}$ 
terms ($\lambda '_{ijk}$). As an example, we consider a model in which only 
$\lambda '_{121}$ and $\lambda '_{221}$ are non-zero, so $\mu^- \rightarrow e^-$ conversion 
is generated at tree-level while $\mu^+ \rightarrow e^+ \gamma$ and $\mu^+ 
\rightarrow e^+e^-e^+$ are generated at one-loop level (Fig.~\ref{diagram3}).

The LFV vertices are
\begin{eqnarray}
D^d &=& -\frac{f^{'}_{121} \lambda '_{221}}{2 m^2_{\tilde{c}_L}},~~~
A_2^R = -\frac{f^{'}_{121} \lambda '_{221}}{64 \pi^2 m^2_{\tilde{d}_R}},
\\
A_1^L &=& \frac{f^{'}_{121} \lambda '_{221}}{96 \pi^2 m^2_{\tilde{d}_R}}
\left\{-5-4 \log\frac{m^2_c}{m^2_{\tilde{d}_R}}-\frac{2}{3}
\delta(m^2_c/q^2)
\right.\nonumber \\
&&\left.+\frac{m^2_{\tilde{d}_R}}{m^2_{\tilde{c}_L}}
\left(-2-2\log \frac{m^2_d}{m^2_{\tilde{c}_L}}-\frac{1}{3} 
\delta(m^2_d/q^2) \right) \right\}.
\end{eqnarray}
As before, we set $q^2=-m^2_\mu$ for $\mu^- \rightarrow e^-$ conversion, and
$q^2=0$ for $\mu^+ \rightarrow e^+e^-e^+$. The ratios of branching ratios are
\begin{eqnarray}
\frac{{\rm Br}(\mu^+ \rightarrow e^+ \gamma)}{{\rm Br}(\mu^+ \rightarrow e^+e^-e^+)}
&=& 1.1,\\
\frac{{\rm R}(\mu^- \rightarrow e^-~{\rm in ~Ti~(Al)})}
{{\rm Br}(\mu^+ \rightarrow e^+e^-e^+)} &=& 2~(1)\times 10^5.
\end{eqnarray}
Here we assume $m_{\tilde{d}_R}=m_{\tilde{c}_L}=300~{\rm GeV}$.
Since $\mu^- \rightarrow e^-$ conversion is induced at the tree-level,
its event rate is much larger than that of other processes, as expected.

In $\mu^+ \rightarrow e^+e^-e^+$, the off-shell photon penguin vertex
($A_1^L$) dominates over the other contributions because of the ultraviolet
log-enhancement. Therefore, the ratio of branching ratios 
${\rm Br}(\mu^+ \rightarrow e^+ \gamma)/{\rm Br}(\mu^+ \rightarrow e^+e^-e^+)$
and the P-odd asymmetries $A_{P_1}$ and $A_{P_2}$ (which are presented in 
Table~\ref{table1}) are very similar to those we obtained in the previous subsection. 
The order one numerical differences come from the different sfermion masses used in
both cases and the fact that there are quarks and not leptons running around the 
loops. Results for other similar examples are also listed in Table~\ref{table1}.

In the case of $\mu^+ \rightarrow e^+e^-e^+$, the fact that we 
choose a fixed value of $q^2(=0)$ instead of integrating over all possible $q^2$ 
values leads to some uncertainty. These, however, are not important as far as our
intentions here are concerned. We note that the numbers presented
in Table~\ref{table1} for ratios of branching ratios when there are first generation 
quarks running around the loops are uncertain by some tens of percent.  

\subsection{Large T-odd Asymmetry in $\mu^+\rightarrow e^+e^-e^+$}

It is important to understand if any interesting effect can be obtained if more
than a single pair of RPV couplings is present. Here we consider the possibility that
$\mu^+\rightarrow e^+e^-e^+$ is generated at the tree-level, but that the loop-level
contributions of on-shell (and off-shell) photons is comparable. This can be
accomplished by having, for example, nonzero $\lambda_{131}\lambda_{231}$ and 
$\lambda_{133}\lambda_{233}\gg \lambda_{131}\lambda_{231}$.  

In this case, all of $B^L, A^L_{1}$, and $A^R_2$ can be comparable, and there is 
the possibility 
that the T-odd asymmetry in $\mu^+\rightarrow e^+e^-e^+$ decay (Eq.(\ref{t-odd}))
is large. We proceed to discuss this in more detail.

We will consider the most general case in which all effective couplings $B^L$,
$A_1^L$, and $A_2^R$ are independent (as may be effectively the case if many
RPV couplings are relevant). In this case, the T-odd asymmetry (Eq.(\ref{t-odd}))
can be written as 
\begin{eqnarray}
A_T &=& \frac{3 (a_{11} C_{11}-a_{12} C_{12})}
{2 (a_2 C_2+a_4 C_4 + a_5 C_5 + a_7 C_7 +a_9 C_9)},
\nonumber \\
&=&\frac{3a_{11}x\sin(\theta_2-\theta_1)-3a_{12}\{y\sin\theta_2
+x\sin(\theta_2-\theta_1)\}}{X},
\end{eqnarray}
where
\begin{eqnarray}
X &=& 4a_2 x^2+4a_4(x^2+y^2+2xy \cos\theta_1)+a_5 
\nonumber \\&&-2a_7x \cos(\theta_2
-\theta_1)-2 a_9 \{y\cos \theta_2+x\cos(\theta_2-\theta_1)\}.
\nonumber
\end{eqnarray}
Here $(a_2,a_4,a_5,a_7,a_9,a_{11},a_{12}) =(1.8,0.96,88,14,7.5,2.0,1,6)$,
$x=|A_1^L/A_2^R|$, $y=|B^L/4\pi \alpha A_2^R|$, and
$\theta_1~(\theta_2)$ is the relative phase between $B^L$ and $A_1^L~(A_2^R)$.
Even when $x,y,\theta_1$ and $\theta_2$ are treated as independent parameters,
this T-odd asymmetry has a maximum value,
\begin{eqnarray}
A_T|_{\rm max} =24\%,
\label{max}
\end{eqnarray}
when
\begin{eqnarray}
x &=& \left|\frac{A_1^L}{A_2^R}\right|=2.56,
\nonumber \\
y &=& \left|\frac{B^L}{4\pi \alpha A_2^R} \right|=4.23, 
\nonumber \\
\theta_1 &=& -2.28,
\nonumber \\
\theta_2 &=& -1.56.
\end{eqnarray}
This upper limit is quite general, and applies to any extension of the SM. It can be 
obtained directly from the most general effective Lagrangian which 
parametrises $\mu^+\rightarrow e^+e^-e^+$ \cite{future}. 
\begin{figure}
\vspace*{-2.0 cm}
\centerline{
\makebox(250,200){\psfig{figure=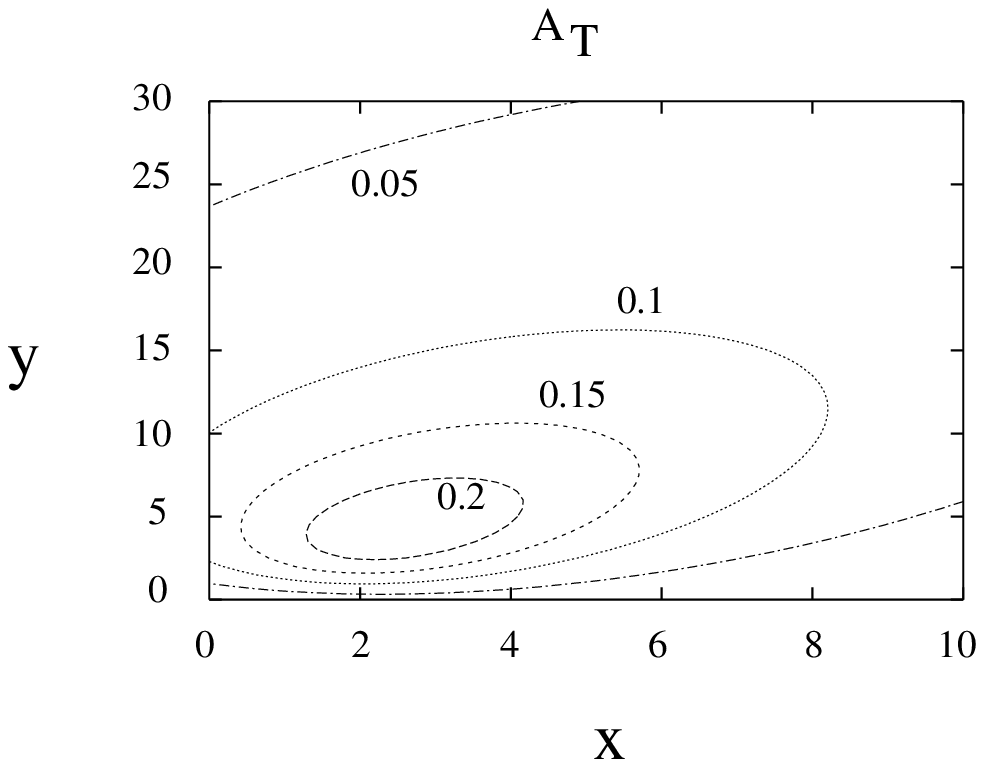,width=0.9\textwidth}}}
\vspace*{1.5 cm}
\centerline{
~~~~~\makebox(250,200){\psfig{figure=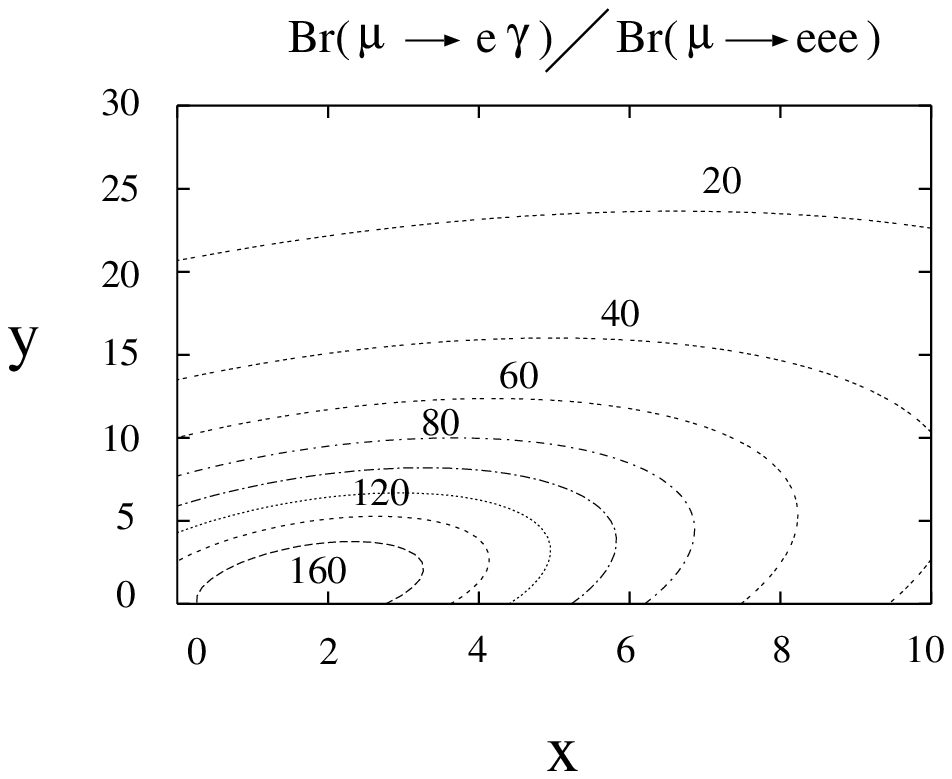,width=0.9\textwidth}}}
\vspace*{1.5 cm}
\caption{\sl Constant contours of the T-odd asymmetry (top) and the ratio of branching 
ratios ${\rm Br}(\mu^+ \rightarrow e^+\gamma)/{\rm Br}(\mu^+ \rightarrow e^+e^-e^+)$ 
(bottom) in the $(x\times y)$~plane. $x=|\frac{A_1^L}{A_2^R}|$ and
$y=|\frac{B^L}{4\pi\alpha A_2^R}|$. The relative phases between 
$B^L,A_2^R$ and $A_1^L, A_2^R$ are fixed at
$(\theta_1, \theta_2)=(-2.28,-1.56)$. See text for details.}
\label{AT}
\end{figure}

Fig.~\ref{AT} depicts the value of the T-odd asymmetry and the ratio of branching 
ratios of $\mu^+\rightarrow e^+\gamma$ and $\mu^+\rightarrow e^+e^-e^+$, when we 
fix $\theta_1=-2.28$, $\theta_2 = -1.56$ (same as at the maximum point). As can be 
seen from Fig.~\ref{AT}, these two observables are strongly correlated. In the region 
where the T-odd asymmetry is relatively large, 
the branching ratio for $\mu^+ \rightarrow e^+ \gamma$ tends to be much bigger than 
the one for $\mu^+ \rightarrow e^+e^-e^+$, since an on-shell
photon coupling $A_2^R$ comparable to $A_1^R$ and $B^L$ is required in order to 
obtain a large T-odd asymmetry. In this case, the branching ratio of $\mu^+\rightarrow
e^+e^-e^+$ is dominated by the $A_{2}^R$ coefficient due to the relatively
large collinear infrared logarithm (see Eq.~(\ref{br(mueee)})) and we obtain a ratio
of branching ratios similar to the one obtained for the MSSM with heavy right-handed
neutrinos (Eq.~(\ref{ratioMSSM1})).

In a generic RPV model, the T-odd asymmetry is unlikely to be close to 
its maximum value (Eq.(\ref{max})) because the branching ratio for $\mu^+ \rightarrow
e^+e^-e^+$ is expected to be comparable to (an in some cases even much larger than) 
the one for $\mu^+ \rightarrow e^+ \gamma$, as we argued in the previous subsections. 
It is, however, possible to tune the various parameters in order to achieve
large effects. In other SUSY extensions of the SM, large T-odd asymmetries can also
be obtained in particular regions of parameter space. For example, the authors of
 \cite{Okada} discuss LFV in the case of SUSY grand unified theories, and find
T-odd asymmetries larger than 15\% in some $SU(5)$ models. 

\begin{figure}
\centerline{
\makebox(250,200){\psfig{figure=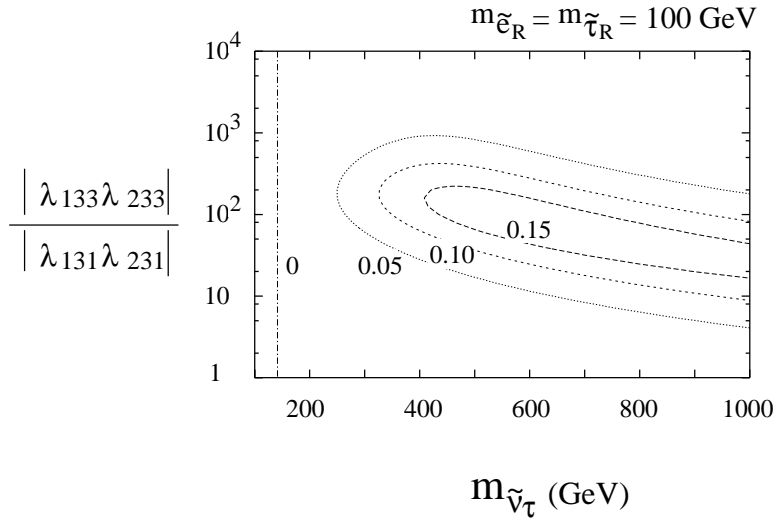,width=0.9\textwidth}}}
\vspace*{0.5 cm}
\caption{\sl Constant contours of the T-odd asymmetry and the in the 
$(m_{\tilde{\nu}_{\tau}}\times |\lambda_{133}\lambda_{233}/
\lambda_{131}\lambda_{231}|)$~plane, assuming
that all other RPV couplings vanishes and that the relative phase between the
two pairs of copings is $\pi/2$.}
\label{large_T-odd}
\end{figure}
As an example, consider a situation where 
$\lambda_{131}\lambda^*_{213}=10^{-6}$ and $\lambda_{133}\lambda^*_{233}=1.6\times 
10^{-4}e^{i\frac{\pi}{2}}$, while all other RPV couplings are zero, leading to
$A_T=17\%$ for $m_{\tilde{\nu}_{\tau}}=500$~GeV and $m_{\tilde{e}_R}=
m_{\tilde{\tau}_R}=100$~GeV. Here Br~$(\mu^+\rightarrow e^+\gamma)=5
\times 10^{-12}$ and Br~$(\mu^+\rightarrow e^+e^-e^+)=3\times 10^{-14}$. Note that, 
in order to obtain large $A_T$ values, one is required to impose a mild hierarchy in 
the ratios of scalar masses (order $10^{1}$) and a more
severe, finely-tuned hierarchy in the ratio of couplings (order $10^{2}$), as
is illustrated in Fig.~\ref{large_T-odd}.

\setcounter{equation}{0}
\section{Conclusions}

We discussed lepton flavour violation (LFV) in rare muon processes 
($\mu^+\rightarrow e^+\gamma$, $\mu^+\rightarrow e^+e^-e^+$, $\mu^-\rightarrow e^-$
conversion in nuclei) in SUSY models with trilinear R-parity violation (RPV). 
Such models are interesting in the sense that they can accommodate neutrino masses 
without requiring the introduction of extra fields to the MSSM. Natural 
explanations for the smallness of the RPV couplings have been studied \cite{MODELS}, 
and are not discussed here.
 
It is well known that LFV in the charged lepton sector is a very sensitive probe for
models with RPV, and that some of the most stringent constraints on RPV couplings 
come from LFV processes. Here, instead of concentrating on how RPV couplings are
constrained by LFV, we study the expectations for LFV observables
in the case nature realizes SUSY with small RPV, and discuss a number of different
observables which may play a decisive role in distinguishing RPV models among 
themselves and from other SUSY models. 

Along these lines, we considered a number of representative cases for different 
RPV models
in order to understand a number of features related to LFV. An important observation
is that, in generic RPV models, all of the LFV processes considered are of the 
same order ({\it i.e.}\/ the ratio of branching ratios is
of order one), {\sl or} $\mu^+\rightarrow e^+\gamma$ is very suppressed with
respect to either $\mu^+\rightarrow e^+e^-e^+$ and/or $\mu^-\rightarrow e^-$
conversion in nuclei, as is summarised in  Table~\ref{table1}. This behaviour is 
to be compared with R-conserving SUSY models with heavy
right-handed neutrinos, 
where 
the branching ratio of $\mu^+\rightarrow e^+\gamma$ is always much larger 
than the branching ratio for $\mu^+\rightarrow e^+e^-e^+$ and (in general)
the rate for $\mu^-\rightarrow e^-$ conversion in nuclei.

We also argue that the P-odd and T-odd asymmetries which can be measured in the case
of polarised $\mu^+\rightarrow e^+e^-e^+$ decays give an extra handle when it 
comes to distinguishing different models. In particular, we discussed whether
a large T-odd asymmetry can be generated in the case of RPV SUSY.

In summary, if there is indeed low-energy SUSY with small but non-negligible
RPV couplings, it is likely that these not only contribute to Majorana neutrino
masses but also will be probed by LFV in the charged lepton sector. If this is the
case, naively higher order QED processes, such as 
$\mu^+\rightarrow e^+e^-e^+$ or $\mu^-\rightarrow e^-$ conversion in nuclei are at 
least as relevant as the more canonical $\mu^+\rightarrow e^+\gamma$ decay. 

Independently of what the new physics beyond the SM is, it should be kept in mind
that improving the current experimental sensitivity of {\sl all} LFV processes is
important. We hope to discuss this important issue in a future publication 
\cite{future}. We conclude by stressing that there are proposals for improving the
sensitivity to $\mu^+\rightarrow e^+\gamma$ down to branchings ratios of $10^{-14}$
\cite{PSI}
and the sensitivity to $\mu^-\rightarrow e^-$ conversion in nuclei down to rates
of $10^{-16}$ \cite{MECO} (see Appendix~B); however, 
in the case of the $\mu^+\rightarrow e^+e^-e^+$, there are no 
proposals for improving the current best bound, which is already twelve years old!
In view of the results discussed here, we believe that experiments which are
sensitive to smaller branching ratios for $\mu^+\rightarrow e^+e^-e^+$ (at least as
sensitive as the future $\mu^+\rightarrow e^+\gamma$ experiments) are of the 
utmost importance.  

\section*{Acknowledgements}
We are indebted to Gian Giudice for many enlightening
discussions and also thank him for carefully
reading the manuscript and providing useful comments.
We acknowledge the Stopped Muon Working Group for 
a Future Muon Storage Ring at CERN for interesting discussions. 
We also thank T.~Yanagida for useful comments. 
      
\appendix
\section{LFV effective vertices in trilinear RPV}
\setcounter{equation}{0}

In this Appendix, we present explicit expressions for the LFV effective
vertices $A_{1,2}^{L,R}$, $B^{L,R}$, and $D^{u,d}$ in the trilinear 
RPV models considered in the body of this paper.

\subsection{Photon penguin vertices}
The photon penguin vertices are defined in Eq.(\ref{photon_penguin}). 
The effective couplings $A_i^{L(R)}$ ($i=1,2$) are given by
\begin{eqnarray}
A_i^{R(L)} &=&A_i^{R(L)(e)}+A_i^{R(L)(\nu)}+A_i^{R(L)(u)}+A_i^{R(L)(d)},
\end{eqnarray}
where $A_i^{R(L)(e,\nu)}$ are induced by R-parity violating
$LL\bar{E}$ couplings through a lepton--sneutrino loop and a neutrino--slepton loop, 
respectively. $A_i^{R(L)(u,d)}$ are generated by $LQ\bar{D}$ couplings through an 
up-type quark--down-type squark loop
and a down-type quark--up-type squark loop, respectively.
The explicit expressions for the on-shell photon vertices $A_2$ are as follows:
\begin{eqnarray}
A_2^{R (e)} &=& 
-\frac{\lambda_{13j} \lambda_{23j}^*}{16 \pi^2 m^2_{\tilde{\nu}3}} 
J_\sigma^{(1)} \left(\frac{m^2_{{e}_{j}}}{m^2_{\tilde{\nu}3}},
\frac{q^2}{m^2_{\tilde{\nu}3}},\frac{m^2_\mu}{m^2_{\tilde{\nu}3}}\right),
\\
A_2^{R (\nu)} &=& \frac{\lambda_{13j} \lambda_{23j}^*}{16 \pi^2 m^2_{\tilde{e}_{Rj}}}
J_\sigma^{(2)}\left(\frac{m^2_{\nu_{3}}}{m^2_{\tilde{e}_{Rj}}},
\frac{q^2}{m^2_{\tilde{e}_{Rj}}},\frac{m^2_\mu}{m^2_{\tilde{e}_{Rj}}}\right),
\\
A_2^{L (e)} &=& 
-\frac{\lambda^*_{ij1} \lambda_{ij2}}{16 \pi^2 m^2_{\tilde{\nu}i}} 
J_\sigma^{(1)} \left(\frac{m^2_{{e}_{j}}}{m^2_{\tilde{\nu}i}},
\frac{q^2}{m^2_{\tilde{\nu}i}},\frac{m^2_\mu}{m^2_{\tilde{\nu}i}}\right),
\\
A_2^{L (\nu)} &=& \frac{\lambda^*_{ij1} \lambda_{ij2}}{16 \pi^2 m^2_{\tilde{e}_{Lj}}}
J_\sigma^{(2)}\left(\frac{m^2_{\nu_{i}}}{m^2_{\tilde{e}_{Lj}}},
\frac{q^2}{m^2_{\tilde{e}_{Lj}}},\frac{m^2_\mu}{m^2_{\tilde{e}_{Lj}}}\right),
\\
A_2^{R (u)} &=& -\frac{\lambda^{'}_{1ij} \lambda^{'*}_{2ij}}
{16 \pi^2 m^2_{\tilde{d}_{Rj}}} \left\{2 
J_\sigma^{(1)}\left(\frac{m^2_{{u}_{i}}}{m^2_{\tilde{d}_{Rj}}},
\frac{q^2}{m^2_{\tilde{d}_{Rj}}},\frac{m^2_\mu}{m^2_{\tilde{d}_{Rj}}}\right)
\right. \nonumber \\
&&\left.\hspace{2.5cm}-J_\sigma^{(2)}\left(\frac{m^2_{{u}_{i}}}
{m^2_{\tilde{d}_{Rj}}},
\frac{q^2}{m^2_{\tilde{d}_{Rj}}},\frac{m^2_\mu}{m^2_{\tilde{d}_{Rj}}}\right)
\right\},
\\
A_2^{R (d)} &=& -\frac{\lambda^{'}_{1ij} \lambda^{'*}_{2ij}}
{16 \pi^2 m^2_{\tilde{u}_{Li}}}
\left\{J_\sigma^{(1)}\left(\frac{m^2_{{d}_{j}}}{m^2_{\tilde{u}_{Li}}},
\frac{q^2}{m^2_{\tilde{u}_{Li}}},\frac{m^2_\mu}{m^2_{\tilde{u}_{Li}}}\right)
\right. \nonumber \\
&&\left.\hspace{2.5cm}-2J_\sigma^{(2)}\left(\frac{m^2_{{d}_{j}}}
{m^2_{\tilde{u}_{Li}}},
\frac{q^2}{m^2_{\tilde{u}_{Li}}},\frac{m^2_\mu}{m^2_{\tilde{u}_{Li}}}\right)
\right\},
\\
A_2^{L (u)}&=&A_2^{L (d)}=0.
\end{eqnarray}
Here the functions $J_\sigma^{(1,2)}$ are defined by
\begin{eqnarray}
&&J_\sigma^{(1)} (a,b,c)
\nonumber \\
&=& \int_0^1dy \int_0^{1-y}dx
\frac{x(1-x-y)}{x\{1-c(1-x-y)\}+a(1-x)-by(1-x-y)},
\\
&&J_\sigma^{(2)} (a,b,c)
\nonumber \\
&=& \int_0^1dy \int_0^{1-y}dx
\frac{y(1-x-y)}{x+y-cy(1-x-y)+a(1-x-y)-bxy}.
\end{eqnarray}
When $b,c\ll 1$, these functions can be approximated by
\begin{eqnarray}
J_{\sigma}^{(1)}(a,b,c) &=& \frac{2+3a-6a^2+a^3+6a \log a}{12(1-a)^4},
\\
J_{\sigma}^{(2)}(a,b,c) &=& \frac{1-6a+3a^2+2a^3-6a^2 \log a}{12(1-a)^4}.
\end{eqnarray}

The off-shell photon vertices $A_1$ are expressed as follows:
\begin{eqnarray}
A_1^{L (e)} &=& \frac{\lambda_{13j} \lambda^*_{23j}}{16 \pi^2 m^2_{\tilde{\nu}3}}
J_q^{(1)}\left(\frac{m^2_{{e}_{j}}}{m^2_{\tilde{\nu}3}},
\frac{q^2}{m^2_{\tilde{\nu}3}},\frac{m^2_\mu}{m^2_{\tilde{\nu}3}}\right),
\\
A_1^{L (\nu)} &=& -\frac{\lambda_{13j} \lambda^*_{23j}}{16 \pi^2 m^2_{\tilde{e}_{Rj}}}
J_q^{(2)}\left(\frac{m^2_{{\nu}_{3}}}{m^2_{\tilde{e}_{Rj}}},
\frac{q^2}{m^2_{\tilde{e}_{Rj}}},\frac{m^2_\mu}{m^2_{\tilde{e}_{Rj}}}\right),
\\
A_1^{R (e)} &=& \frac{\lambda^*_{ij1} \lambda_{ij2}}{16 \pi^2 m^2_{\tilde{\nu}i}}
J_q^{(1)}\left(\frac{m^2_{{e}_{j}}}{m^2_{\tilde{\nu}i}},
\frac{q^2}{m^2_{\tilde{\nu}i}},\frac{m^2_\mu}{m^2_{\tilde{\nu}i}}\right),
\\
A_1^{R (\nu)} &=& -\frac{\lambda^*_{ij1} \lambda_{ij2}}{16 \pi^2 m^2_{\tilde{e}_{Lj}}}
J_q^{(2)}\left(\frac{m^2_{{\nu}_{i}}}{m^2_{\tilde{e}_{Lj}}},
\frac{q^2}{m^2_{\tilde{e}_{Lj}}},\frac{m^2_\mu}{m^2_{\tilde{\nu}i}}\right),
\\
A_1^{L (u)} &=& \frac{\lambda^{'}_{1ij} \lambda^{'*}_{2ij}}{16 
\pi^2 m^2_{\tilde{d}_{Rj}}}
\left\{2 J^{(1)}_q \left(\frac{m^2_{u_i}}{m^2_{\tilde{d}_{Rj}}},
\frac{q^2}{m^2_{\tilde{d}_{Rj}}},\frac{m^2_\mu}{m^2_{\tilde{d}_{Rj}}}
\right)\right. \nonumber \\
&&\left.\hspace{2.5cm}-J^{(2)}_q \left(\frac{m^2_{u_i}}{m^2_{\tilde{d}_{Rj}}},
\frac{q^2}{m^2_{\tilde{d}_{Rj}}},\frac{m^2_\mu}{m^2_{\tilde{d}_{Rj}}}
\right)\right\},
\\
A_1^{(d)} &=& \frac{\lambda^{'}_{1ij} \lambda^{'*}_{2ij}}{16 
\pi^2 m^2_{\tilde{u}_{Li}}}
\left\{J^{(1)}_q \left(\frac{m^2_{d_j}}{m^2_{\tilde{u}_{Li}}},
\frac{q^2}{m^2_{\tilde{u}_{Li}}},\frac{m^2_\mu}{m^2_{\tilde{u}_{Li}}}
\right)\right. \nonumber \\
&&\left.\hspace{2.5cm}-2J^{(2)}_q \left(\frac{m^2_{d_j}}{m^2_{\tilde{u}_{Li}}},
\frac{q^2}{m^2_{\tilde{u}_{Li}}},\frac{m^2_\mu}{m^2_{\tilde{u}_{Li}}}
\right)\right\},
\end{eqnarray}
where the functions $J_q^{(1,2)}$ are defined by
\begin{eqnarray}
&&J_q^{(1)}(a,b,c)
\nonumber \\
&=& \int_0^1 dy \int_0^{1-y}dx
\frac{(x+2y)(1-x-y)}{x\{1-c(1-x-y)\}+a(1-x)-by(1-x-y)},
\\
&&J_q^{(2)}(a,b,c)
\nonumber \\
&=& \int_0^1 dy \int_0^{1-y}dx
\frac{y(x-y)}{x+y-cy(1-x-y)
+a(1-x-y)-bxy}.
\end{eqnarray}
When $a,b,c\ll 1$, these functions are well approximated by
\begin{eqnarray}
J_{q}^{(1)}(a,b,c) &=& -\frac{1}{3}
\left(\frac{4}{3}+\log{a}+\delta (a/b)
\right),
\\
J_q^{(2)}(a,b,c) &=& -\frac{1}{18},
\end{eqnarray}
where
\begin{eqnarray}
\delta(d) &=&
\left\{
\begin{array}{l}
-\frac{5}{3}-4d+2(1+2d)\sqrt{1-4d}\tanh^{-1}
\frac{1}{\sqrt{1-4d}}, ~~{\rm for}~d<\frac{1}{4},
\\
-\frac{5}{3}-4d+2(1+2d)\sqrt{4d-1}\tan^{-1}
\frac{1}{\sqrt{4d-1}}, ~~{\rm for}~d>\frac{1}{4}.
\end{array}
\right.
\end{eqnarray}
When $a \gg b,c$ and $b,c\ll 1$,
\begin{eqnarray}
J_q^{(1)} (a,b,c) &=& \frac{-16+45a-36a^2+7a^3-(12-18a)\log a}
{36 (1-a)^4},
\\
J_q^{(2)} (a,b,c) &=& \frac{-2+9a-18a^2+11a^3-6a^3 \log a}{36(1-a)^4}.
\end{eqnarray}

\subsection{Tree-level vertices in the $\mu^+ \rightarrow e^+e^-e^+$ process}

The tree-level vertices $B^{R(L)}$ in the $\mu^+ \rightarrow e^+e^-e^+$ process
are defined in Eq.(\ref{B_vertex}). Their explicit expressions are given by
\begin{eqnarray}
B^L &=& -\frac{\lambda_{i11}\lambda^*_{i21}}{2 m^2_{\tilde{\nu}i}},
\\
B^R &=& -\frac{\lambda^*_{i11}\lambda_{i12}}{2 m^2_{\tilde{\nu}i}}.
\end{eqnarray}

\subsection{Tree-level vertices in the $\mu^- \rightarrow e^-$ conversion process}

The tree-level vertices for $\mu^- \rightarrow e^-$ conversion were defined in 
Eq.(\ref{D_vertex}). Their explicit forms are 
\begin{eqnarray}
D^{u} &=& \frac{\lambda^{'*}_{11i} \lambda^{'}_{21i}}{2 m^2_{\tilde{d}_{Ri}}},
\\
D^{d} &=& -\frac{\lambda^{'*}_{1j1} \lambda^{'}_{2j1}}{2 m^2_{\tilde{u}_{Lj}}}.
\end{eqnarray}

\section{Constraints on R-parity-violating Couplings from LFV Processes and
Neutrino Masses}

The experimental limits on LFV processes set tight bounds on specific 
combinations of R-parity violating couplings. The most stringent experimental limit 
on the branching ratio of $\mu^+ \rightarrow e^+e^-e^+$ is given by the SINDRUM 
experiment at PSI~\cite{mu3e}:\footnote{In order to reach the current bounds,
rare muon decay experiments need to stop the muons before they decay. For this reason,
they are constrained to analyse $\mu^+$ decays, since the $\mu^-$ is readily captured 
by the material present in order to stop the muons and there are virtually no 
free $\mu^-$ decays. For the same reason, one can only measure the $\mu\rightarrow e$ 
conversion rate in nuclei for the $\mu^-$.}
\begin{eqnarray}
{\rm Br}(\mu^+ \rightarrow e^+e^-e^+)|_{\rm exp.} < 1.0 \times 10^{-12}.
\end{eqnarray}
The present experimental limit on the branching ratio of 
$\mu^+ \rightarrow e^+ \gamma$ process is set by the MEGA collaboration
at LANL~{\cite{MEGA}}:
\begin{eqnarray}
{\rm Br}(\mu^+ \rightarrow e^+ \gamma)|_{\rm exp.}<1.2 \times 10^{-11}.
\end{eqnarray}
This limit will be significantly improved (or, perhaps, LFV will be found!)
in the near future by a new experiment at PSI~\cite{PSI}, which claims to be able to
observe $\mu^+\rightarrow e^+ \gamma$ events if  
${\rm Br}(\mu^+ \rightarrow e^+ \gamma)> 10^{-14}$.
The present experimental bound on the conversion rate of $\mu^- \rightarrow e^-$ 
in $^{48}_{22}{\rm Ti}$ was determined by the SINDRUM 2 collaboration
at PSI~{\cite{mueconv}}:
\begin{eqnarray}
{\rm R}(\mu^- \rightarrow e^- ~{\rm in}~{^{48}_{22}{\rm Ti}})|_{\rm exp.}
<6.1 \times 10^{-13}.
\end{eqnarray}
The future proposed (almost approved) experiment MECO \cite{MECO} 
claims that it will be able 
to see $\mu^-\rightarrow e^-$ conversion in aluminium if ${\rm R}(\mu^- \rightarrow 
e^-~{\rm in}~{^{27}_{13}{\rm Al}}) > 10^{-16}$. (More futuristic proposals 
claim sensitivity to values of the rate of $\mu^-\rightarrow e^-$ conversion in nuclei 
as low as $10^{-18}$ \cite{PRISM}!)  

Tables~\ref{table2}, \ref{table3} contain current and near future bounds on
the absolute values of some pairs of RPV couplings, assuming that all other
pairs of couplings vanish.   

\begin{table}
\caption{\sl Current (future) constraints on the R-parity violating 
couplings $LL\bar{E}$ (see Eq.~(\ref{wrpv})) from LFV processes, assuming that only 
the listed pair of coupling is nonzero. The current (future) upper limits  
on the branching ratios are: 
${\rm Br}(\mu^+ \rightarrow e^+\gamma)<1.2\times10^{-11}~(10^{-14})$,
${\rm Br}(\mu^+ \rightarrow e^+e^-e^+)<1.0\times 10^{-12}$, and
${\rm R}(\mu^- \rightarrow e^-~{\rm in~Ti}) < 6.1\times 10^{-13}$
(${\rm R}(\mu^- \rightarrow e^-~{\rm in~Al}) < 10^{-16}$).Here we assume all
the sneutrino masses degenerate with right-handed slepton masses, 
$m_{\tilde{\nu},\tilde{l}_R}=100$~GeV,
and we neglect left-right mixing in the charged slepton mass matrix. 
The notation (tree) indicates that the $\mu^+ \rightarrow e^+e^-e^+$ process 
is generated at the tree-level.}
\begin{center}
\begin{tabular}{|c|c|c|c|} \hline
 & $\mu \rightarrow e\gamma$ & $\mu \rightarrow eee$ 
& $\mu \rightarrow e~{\rm in~nuclei}$
\\ \hline
$|\lambda_{131}\lambda_{231}|$ & $2.3\times 10^{-4}$~\cite{Chaichian}
&$6.7\times 10^{-7}({\rm tree})$~
\cite{Choudhury} &$1.1\times 10^{-5}$~\cite{Huitu}
\\&$(7\times 10^{-6})$ && $(2\times 10^{-7})$
\\
$|\lambda_{132}\lambda_{232}|$ & $2.3\times 10^{-4}$~\cite{Chaichian}
& $7.1\times 10^{-5}$
 &$1.3\times 10^{-5}$~\cite{Huitu} 
\\
& $(7\times 10^{-6})$ & & $(2\times 10^{-7})$
\\
$|\lambda_{133}\lambda_{233}|$ & $2.3\times 10^{-4}$
~\cite{Chaichian} & $1.2\times 10^{-4}$
 & $2.3\times 10^{-5}$~\cite{Huitu} 
\\
 & $(7\times 10^{-6})$ &  & $(4\times 10^{-7})$
\\
$|\lambda_{121}\lambda_{122}|$ & $8.2\times 10^{-5}$ ~\cite{Chaichian}
&$6.7\times 10^{-7}({\rm tree})$
~\cite{Choudhury}
&$ 6.1\times 10^{-6}$~\cite{Huitu}
\\
 & $(2\times 10^{-6})$ &  & $(1\times 10^{-7})$
\\
$|\lambda_{131}\lambda_{132}|$ & $8.2\times 10^{-5}$ 
~\cite{Chaichian} &$6.7\times 10^{-7}({\rm tree})$
~\cite{Choudhury}
&$7.6 \times 10^{-6}$~\cite{Huitu}
\\
 & $(2\times 10^{-6})$ &  & $(1\times 10^{-7})$
\\
$|\lambda_{231}\lambda_{232}|$ & $8.2\times 10^{-5}$
~\cite{Chaichian} & $4.5\times 10^{-5}$ &
$8.3\times 10^{-6}$~\cite{Huitu}
\\
 & $(2\times 10^{-6})$ &  & $(1\times 10^{-7})$
\\ \hline
\end{tabular}
\end{center}
\label{table2}
\end{table}

\begin{table}
\caption{\sl Current (future) constraints on the R-parity violating 
couplings $LQ\bar{D}$ (see Eq.~(\ref{wrpv})) from LFV processes, assuming that only 
the listed pair of coupling is nonzero. The current (future) upper limits  
on the branching ratios are: 
${\rm Br}(\mu^+ \rightarrow e^+\gamma)<1.2\times10^{-11}~(10^{-14})$,
${\rm Br}(\mu^+ \rightarrow e^+e^-e^+)<1.0\times 10^{-12}$, and
${\rm R}(\mu^- \rightarrow e^-~{\rm in~Ti}) < 6.1\times 10^{-13}$
(${\rm R}(\mu^- \rightarrow e^-~{\rm in~Al}) < 10^{-16}$). Here we assume all
the squark masses are degenerate, with $m_{\tilde{q}}=300$~GeV. The notation 
(tree) indicates that
the $\mu^- \rightarrow e^-$ conversion process is generated at the tree-level.}
\begin{center}
\begin{tabular}{|c|c|c|c|} \hline
 & $\mu \rightarrow e\gamma$ & $\mu \rightarrow eee$ 
& $\mu \rightarrow e~{\rm in~nuclei}$
\\ \hline
$|\lambda '_{111}\lambda '_{211}|$ & $6.8\times 10^{-4}$~\cite{Chaichian}
& $1.3\times 10^{-4}$ &
$5.4\times 10^{-6}~({\rm tree})$~\cite{Kim}
\\
 & $(2\times 10^{-5})$ & & $(2\times 10^{-7})$
\\
$|\lambda '_{112}\lambda '_{212}|$ & $6.8\times 10^{-4}$~\cite{Chaichian} 
& $1.4\times 10^{-4}$ &
$3.9\times 10^{-7}~({\rm tree})$~\cite{Kim}
\\
 & $(2\times 10^{-5})$ & & $(7\times 10^{-9})$
\\
$|\lambda '_{113}\lambda '_{213}|$ & $6.8\times 10^{-4}$~\cite{Chaichian}
& $1.6\times 10^{-4}$ &
$3.9\times 10^{-7}~({\rm tree})$~\cite{Kim}
\\
 & $(2\times 10^{-5})$ & & $(7\times 10^{-9})$
\\
$|\lambda '_{121}\lambda '_{221}|$ & $6.8\times 10^{-4}$~\cite{Chaichian}
& $2.0\times 10^{-4}$ &
$3.6\times 10^{-7}~({\rm tree})$~\cite{Kim}
\\
 & $(2\times 10^{-5})$ & & $(6\times 10^{-9})$
\\
$|\lambda '_{122}\lambda '_{222}|$ & $6.8\times 10^{-4}$~\cite{Chaichian}
& $2.3\times 10^{-4}$ &
$4.3\times 10^{-5}$~\cite{Huitu}
\\
 & $(2\times 10^{-5})$ & & $(7\times 10^{-7})$ 
\\
$|\lambda '_{123}\lambda '_{223}|$ & $6.9\times 10^{-4}$~\cite{Chaichian}
& $2.9\times 10^{-4}$ &
$5.4\times 10^{-5}$~\cite{Huitu}
\\
 & $(2\times 10^{-5})$ & & $(9\times 10^{-7})$ 
\\ \hline
\end{tabular}
\end{center}
\label{table3}
\end{table}

In models with trilinear RPV, neutrino masses are generated 
at one-loop via squark (slepton) exchange for $LQ\bar{D}$ ($LL\bar{E}$) operators.
Under the assumption that the left-right sfermion soft mass-squared mixing terms 
are diagonal in the physical basis and proportional to the associated fermion mass
($m^{2}_{\tilde{f}LR}\propto m_fm_{\tilde{f}}$),
the formula for the neutrino masses can be simplified to \cite{bhat}
\begin{eqnarray} 
m_{\nu_{ii'}} & \simeq & {{n_c \lambda_{ijk} \lambda_{ikj}}
\over{16\pi^2}} m_{f_j} m_{f_k}
\left [ \frac{f(m^2_{f_j}/m^2_{\tilde{f}_k})} {m_{\tilde{f}_k}} +
\frac{f(m^2_{f_k}/m^2_{\tilde{f}_j})} {m_{\tilde{f}_j}}\right ]  \\
f(x) & = & (x\ln x-x+1)/(x-1)^2 \nonumber
\end{eqnarray}  
Here, $m_{f_i}$ is the fermion
mass of the $i$th generation inside the loop, $m_{\tilde{f}_i}$ is the
average of the $\tilde{f}_{Li}$ and $\tilde{f}_{Ri}$ squark masses, and
$n_c$ is a colour factor (3 for
$LQ\bar{D}$ operators and 1 for $LL\bar{E}$ operators). This expression implies that 
the heavier the fermions in the loop, the stricter the bounds \cite{bhat}. 
For example, demanding $m_{e\mu}<1$ eV for sparticle masses of 300 GeV, 
$m_b = 4.4$~GeV and $m_s = 170$ MeV, leads to $\lambda^{'}_{133}\lambda^{'}_{233} 
\leq 4 \cdot 10^{-7}$. For $\lambda^{'}_{122}\lambda^{'}_{222}$  the bound
drops to  $2.3 \cdot 10^{-4}$ \cite{bhat}, while for ``Super-Kamiokande-friendly'' 
solutions with hierarchical neutrinos the bounds on certain products of RPV 
couplings can be stricter by some orders of magnitude. 

When comparing these bounds with the ones from LFV in Tables~\ref{table2} and 
\ref{table3}, we see that for a large number
of models the bounds from stopped muon processes
are significantly stronger than those from neutrino masses. A proper study of these 
processes therefore, can shed additional light in the issue of lepton number violation.

%
%
%
\newcommand{\Journal}[4]{{\sl #1} {\bf #2} {(#3)} {#4}}
\newcommand{\APJ}{Ap. J.}
\newcommand{\CJP}{Can. J. Phys.}
\newcommand{\NC}{Nuovo Cimento}
\newcommand{\NP}{Nucl. Phys.}
\newcommand{\PL}{Phys. Lett.}
\newcommand{\PR}{Phys. Rev.}
\newcommand{\PRep}{Phys. Rep.}
\newcommand{\PRL}{Phys. Rev. Lett.}
\newcommand{\PTP}{Prog. Theor. Phys.}
\newcommand{\SJNP}{Sov. J. Nucl. Phys.}
\newcommand{\ZP}{Z. Phys.}
\newcommand{\EUR}{Eur. Phys. J.}


\begin{thebibliography}{99}
%
%
\bibitem{atm} Y. Fukuda {\it et al.} (Super-Kamiokande Collaboration),
\Journal{\PRL}{81}{1998}{1562}.
%
\bibitem{solar} B.T.~Cleveland {\it et al.,}\/ {\sl Astrophys.~J.}{\bf 496}(1998) 505;
Dzh.N. Abdurashitov {\it et al.}\/ (SAGE collaboration), 
\Journal{\PRL}{77}{1996}{4708};
W. Hampel{\it  et al.} (GALLEX Collaboration), \Journal{\PL}{B447}{1999}{127};
Y. Fukuda {\it et al.} (Super-Kamiokande Collaboration), 
\Journal{\PRL}{82}{1999}{1810}; {\sl ibid.} {\bf 82} (1999) 2430.
%
\bibitem{LSND} C. Athanassopoulos {\it et al.}\/ (LSND Collaboration),
\Journal{\PRL}{77}{1996}{3082}; {\sl ibid.} {\bf 81} (1998) 1774.
%
\bibitem{seesaw} T. Yanagida, in {\it Proceedings of the Workshop on
Unified Theory and Baryon Number of the Universe}, Tsukuba, Japan,
1979, edited by O. Sawada and A. Sugamoto (KEK, Tsukuba, 1979), p. 95;
M. Gell-Mann, P. Ramond, and R. Slansky, in {\it Supergravity}, Proceedings
of the Workshop, Stony Brook, New York, 1979, edited by P. van Nieuwenhuizen 
and D. Freedman (North-Holland, Amsterdam, 1979).
%
\bibitem{neutrinoLFVns}
M.~Nakagawa, H.~Okonogi, S.~Sakata, and A.~Toyoda, {\sl Prog. Theor. Phys.}\/ {\bf 30}
(1963), 727; S.~Eliezer and D.~Ross, \Journal{\PR}{D10}{1974}{3088}; 
S.M.~Bilenki and B.~Pontecorvo, \Journal{\PL}{61B}{1976}{248}; S.~Barshay, 
\Journal{\PL}{63B}{1976}{466}; 
T.P.~Cheng and L.-F.~Li in {\it Proceedings of the Coral Gables Conference}, 1977,
ed. A.~Perlmutter, Plenum, New York.
A.~Mann and H.~Primakoff, \Journal{\PR}{D15}{1977}{655};
S.T. Petcov, {\sl Yad. Phys.}\/ {\bf 25} (1977) 641 and {\sl Sov. J. Nucl. Phys.}\/ 
{\bf 25} (1977) 340; S.M. Bilenki, S.T. Petcov and B. Pontecorvo,
{\sl Phys. Lett.}\/ {\bf B67} (1977) 309.
%
\bibitem{masiero} J.~Ellis and D.V.~Nanopoulos, {\sl Phys. Lett.}\/ {\bf  B110} 
(1982) 44; R.~Barbieri and R.~Gatto, {\sl Phys. Lett.}\/ {\bf  B110} (1982) 211;
G.K.~Leontaris, K.~Tamvakis and J.D.~Vergados, {\sl Phys. Lett.}\/ {\bf B171}
 (1986) 412; F.~Borzumati, and A.~Masiero, \Journal{\PRL}{57}{1986}{961}.
F.~Gabianni and A.~Masiero, {\sl Nucl. Phys.}\/ {\bf B322} (1989) 235;
%
\bibitem{tobe} J. Hisano, T. Moroi, K. Tobe, M. Yamaguchi, and T. Yanagida,
  \Journal{\PL}{B357}{1995}{579};
  J. Hisano, T. Moroi, K. Tobe, and M. Yamaguchi,
  \Journal{\PR}{D53}{1996}{2442}. K.~Tobe, {\sl Nucl. Phys. Proc. Suppl.}
 {\bf B59} (1997) 223.
%
\bibitem{hisano} S.~Dimopoulos and D.~Sutter, \Journal{NP}{B 452}{1995}{496}; 
B.~de Carlos, J.A.~Casas and J.M.~Moreno, 
\Journal{\PR}{D53}{1996}{6398}; J.~Hisano, D.~Nomura, T.~Yanagida,
  \Journal{\PL}{B437}{1998}{351}; J.~Hisano and D.~Nomura,
  \Journal{\PR}{D59}{1999}{116005};
M. G\'omez, G. Leontaris, S. Lola and J. Vergados, \Journal{\PR}{D59}{1999}{116009};
J.~Ellis, M.E.~Gomez, G.K.~Leontaris, S.~Lola,
and D.V.~Nanopoulos, \Journal{\EUR}{C14}{2000}{319};
W.~Buchmuller, D.~Delepine, and F.~Vissani,
\Journal{\PL}{B459}{1999}{171}; W.~Buchmuller, D.~Delepine,
and L.T.~Handoko, \Journal{\NP}{B576}{2000}{445}; 
J.L.~Feng, Y.~Nir and Y.~Shadmi, \Journal{\PR}{D61}{2000}{113005}
S.~Baek, T.~Goto, Y.~Okada, and
K.~Okumura, hep-ph/0002141.
%
\bibitem{Okada_kuno} For a recent review, see Y.~Kuno and Y.~Okada,
hep-ph/9909265.
%
\bibitem{ROSS} L.E.~Ibanez, and G.G.~Ross,
  \Journal{\NP}{B368}{1992}{3}.
%
\bibitem{Fukugita} M.~Fukugita and T.~Yanagida, 
  \Journal{\PR}{D42}{1990}{1285};
B.A. Campbell, S. Davidson, J. Ellis, and K.A. Olive,
\Journal{\PL}{B256}{1991}{457}; 
W. Fischler, G.F. Giudice, R.G. Leigh, and S. Paban, 
\Journal{\PL}{B258}{1991}{45}.
%
\bibitem{EWBaryon} For a recent review of baryogenesis at the electroweak
scale within supersymmetry see: M. Carena, C.E.M. Wagner, ``{\it 
Perspectives on Higgs Physics II}'', ed. G.L. Kane, World Scientific,
Singapore.
%
\bibitem{neutrino_Rp} L. Hall and M. Suzuki,
\Journal{\NP}{B231}{1984}{419}; A.S. Joshipura and M. Nowakowski,
\Journal{\PR}{D51}{1995}{2421}; 
T. Banks, Y. Grossman, E. Nardi, and Y. Nir, \Journal{\PR}{D52}{1996}{5319};
F.M. Borzumati, Y. Grossman, E. Nardi,
and Y. Nir, \Journal{\PL}{B384}{1996}{123}; B. de Carlos and P.L. White,
\Journal{\PR}{D54}{1996}{3427}; A.Yu. Smirnov and F. Vissani,
\Journal{\NP}{B460}{1996}{37}; R. Hempfling, \Journal{\NP}{B478}{1996}{3};
H.P. Nilles and N. Polonsky, \Journal{\NP}{B484}{1997}{33};
E. Nardi, \Journal{\PR}{D55}{1997}{5772}; 
M. Hirsch, H.V. Klapdor-Kleingrothaus, and S.G. Kovalenko, \Journal{\PR}{D57}
{1998}{1947};
E.J. Chun, S.K. Kang, C.W. Kim, and
U.W. Lee, \Journal{\NP}{B544}{1999}{89}; 
O.C.W. Kong, {\sl Mod. Phys. Lett.} {\bf A14} (1999) 903;
L. Clavelli and P.H. Frampton, hep-ph/9811326; 
S. Rakshit and G. Bhattacharyya, and A. Raychaudhuri;
\Journal{\PR}{D59}{1999}{091701}; R. Adhikari and G. Omanovic, 
\Journal{\PR}{D59}{1999}{073003}; D.E. Kaplan and A.E. Nelson,
{\sl JHEP} {\bf 0001} (2000) 033;
A.S. Joshipura and S.K. Vempati,
\Journal{\PR}{D60}{1999}{095009}; {\sl ibid} {\bf D60} (1999) 111303;
J. Ferrandis, \Journal{\PR}{D60}{1999}{095012}; M. Bisset, O.C.W. Kong,
C. Macesanu, and L.H. Orr, \Journal{\PR}{D62}{2000}{035001}; 
Y. Grossman and H.E. Haber, hep-ph/9906310;
A. Abada and M. Losada, hep-ph/9908352, hep-ph/0007041; 
O. Haug, J.D. Vergados, A. Faessler, and S. Kovalenko,
\Journal{\NP}{B565}{2000}{38};
E.J. Chun and S.K. Kang, \Journal{\PR}{D61}{2000}{075012};
F. Takayama and M. Yamaguchi, \Journal{\PL}{B476}{2000}{116};
R. Kitano and K. Oda, \Journal{\PR}{D61}{2000}{113001}.
S. Davidson and M. Losada, {\sl JHEP} {\bf 0005} (2000) 021;
J.C. Romao, M.A. Diaz, M. Hirsch,
W. Porod, and J.W.F. Valle, \Journal{\PR}{D61}{2000}{071703};
hep-ph/0004115.
%
\bibitem{pila} M. Nowakowski and A. Pilaftsis, \Journal{\NP}{B461}{1996}{19}; 

\bibitem{bhat}
G. Bhattacharyya,
H.V. Klapdor-Kleingrothaus and H. Pas, \Journal{\PL}{B463}{1999}{77}.
%
\bibitem{MODELS}  
S.~Lola and G.G.~Ross, {\sl Phys. Lett.}\/ {\bf B314} (1993) 336;
V.~Ben-Hamo, Y.~Nir, {\sl Phys. Lett.}\/ {\bf B339} (1994) 77; 
H.~Dreiner and A.~Chamseddine, {\sl Nucl. Phys.} {\bf B 458} (1996) 65; 
P.~Binetruy, S.~Lavignac and P.~Ramond, Nucl. Phys. B477 (1996) 353; 
G.~Bhattacharyya, {\sl Phys. Rev.}\/ {\bf D57} (1998) 3944;
P.~Binetruy, E.~Dudas, S.~Lavignac and C.A.~Savoy, 
{\sl Phys. Lett.}\/ {\bf B422} (1998) 171;
J. Ellis, S. Lola and G.G. Ross, {\sl Nucl. Phys.}\/ {\bf B526} (1998) 115.
%
\bibitem{Choudhury} D.~Choudhury and P.~Roy,
  \Journal{\PL}{B378}{1996}{153}.
%
\bibitem{Chaichian} M.~Chaichian, and K.~Huitu,
  \Journal{\PL}{B384}{1996}{157}.
%
\bibitem{Kim} J.E.~Kim, P.~Ko, and D.-G.~Lee,
  \Journal{\PR}{D56}{1997}{100}.
%
\bibitem{Huitu} K.~Huitu, J.~Maalampi, M.~Raidal, and A.~Santamaria,
  \Journal{\PL}{B430}{1998}{355}.
%
\bibitem{Faessler} 
A.~Faessler, T.S.~Kosmas, S.~Kovalenko, and J.D.~Vergados, hep-ph/9904335.
%
\bibitem{Choi}
K.~Choi, E.J.~Chun, K.~Hwang, hep-ph/0005262.
%
\bibitem{HERA} M.~Derrick {\it et al}\/ (ZEUS Collaboration), 
\Journal{\ZP}{C 73}{1997}{613}; C.~Adloff {\it et al.}\/ (H1 Collaboration), 
\Journal{\EUR}{C 11}{1999}{447}, err. {\bf C 14} (2000) 553; R.~Kerger, 
hep-ex/0006023, talk at {\it DIS2000 (8$^{th}$ International Workshop on Deep
Inelastic Scattering and QCD)}, Liverpool, 25--30 April 2000;
M.~Kuze, S.~Lola, E.~Perez and B.C.~Allanach,
hep-ph/0007282, summary of {\it DIS2000}.



\bibitem{CLEO}
S.~Ahmed {\it et al.}, CLEO Collaboration, \Journal{\PR}{D61}{2000}{071101}
%
\bibitem{kuno} Y.~Kuno and Y.~Okada,
  \Journal{\PRL}{77}{1996}{434}; Y.~Kuno, A.~Maki, and Y.~Okada,
\Journal{\PR}{D55}{1997}{2517}.
%
\bibitem{TWZ} S.B.~Treiman, F.~Wilczek, and A.~Zee, \Journal{\PR}{D16}{1977}{152};
A.~Zee, \Journal{\PRL}{55}{1985}{55}.

\bibitem{Okada} Y.~Okada, K.~Okumura, and Y.~Shimizu,
  \Journal{\PR}{D58}{1998}{051901}; {\sl ibid.} {\bf D61} (2000) 094001.
%
\bibitem{capture} T. Suzuki, D.F. Measday, and J.P. Roalsvig,
\Journal{\PR}{C35}{1987}{2212}.
%
\bibitem{mu_e_conv} H.C. Chiang, E. Oset, T.S. Kosmas, A. Faessler,
and J.D. Vergados, \Journal{\NP}{A559}{1993}{526}; 
T.S.~Kosmas, A.~Faessler, F.~Simkovic and J.D.~Vergados, 
\Journal{\PR}{C56}{1997}{526}.
%
\bibitem{Sanda} Similar log-enhancements in other models have been
discussed. For example,
F.~Wilczek and A.~Zee, \Journal{\PRL}{38}{1977}{531};
W.J.~Marciano and A.I.~Sanda, \Journal{\PRL}{38}{1977}{1512};
M.~Raidal and A.~Santamaria, \Journal{\PL}{B421}{1998}{250}.
%
\bibitem{GUTs} L.J.~Hall, V.A.~Kostelecky, and S.~Raby, \Journal{\NP}{B 267}{1986}{415};
A. E. Faraggi, J.L. Lopez, D.V. Nanopoulos and K. Yuan, \Journal{\PL}{B221}{1989}{337};
S.~Kelley, J.L.~Lopez, D.V.~Nanopoulos and H.~Pois, \Journal{\NP}{B 358}{1991}{27}; 
R.~Barbieri and L.J.~Hall, \Journal{\PL}{B338}{1994}{212}; R.~Barbieri, L.J.~Hall, and
A.~Strumia, \Journal{\NP}{B 445}{1995}{219};
A.~Ilakovac and A.~Pilaftsis, \Journal{\NP}{B 437}{1995}{491};
P.~Ciafaloni, A.~Romanino, and A.~Strumia, \Journal{\NP}{B 458}{1996}{3};
N.~Arkani-Hamed, H.-C.~Cheng, and L.J.~Hall, \Journal{\PR}{D53}{1996}{413};
M.E.~G\'omez and H.~Goldberg, \Journal{\PR}{D53}{1996}{5244};
N.G.~Deshpande, B.~Dutta, and E.~Keith, \Journal{\PR}{D54}{1996}{730} 
T.V.~~Duong, B.~Butta, and E.~Keith, \Journal{\PL}{B378}{1996}{128}
J.~Hisano, T.~Moroi, K.~Tobe, and
M.~Yamaguchi, \Journal{\PL}{B391}{1997}{341}, erratum {\bf B397} (1997) 357; 
D.~Suematsu, \Journal{\PL}{B416}{1998}{108}
J.~Hisano, D.~Nomura, Y.~Okada, and M.~Tanaka, \Journal{\PR}{D58}{1998}{116010}; 
S.F.~King and M.~Oliveira, \Journal{\PR}{D60}{1999}{035003};
G.~Couture, M.~Frank, H.~Konig, and M.~Pospolov, {\sl Euro. Phys.}\/ {\bf C7} (1999)
139; K.~Kurosawa and N.~Maekawa, {\sl Prog. Theor. Phys.}\/ {\bf 102} (1999) 121;
Y.~Okada and K.~Okumura, \Journal{\PR}{D61}{2000}{094001}; 
R.~Kitano and K.~Yamamoto, hep-ph/9905459; 
G.~Barenboim, K.~Huitu, and M.~Raidal, hep-ph/0005159. 
%
\bibitem{future} A.~de Gouv\^ea {\it et al.,}\/ in preparation.
%
\bibitem{PSI} L.M.~Barkov {\it et al.,}\/
Research Proposal to PSI, 1999, \verb|http://www.icepp.s.u-tokyo.ac.jp/meg|
%
\bibitem{MECO} M.~Bachman {\it et al.}\/ (MECO Collaboration),
Proposal to BNL, 1997. See also \verb|http://meco.ps.uci.edu|.
%
\bibitem{mu3e} U.~Bellgardt, {\it et al.,}\/ 
  \Journal{\NP}{B229}{1988}{1}.
%
\bibitem{MEGA} M.L.~Brooks, {\it et al.}\/ (MEGA collaboration), 
\Journal{\PRL}{83}{1999}{1521}. 
%
\bibitem{mueconv} P.~Wints (SINDRUM 2 collaboration), 1998
in {\it Proceedings of the First International Symposium
on Lepton and Baryon Number Violation}, ed. H.V.~Klapdor-Kleingrothaus
and I.V.~Krivosheina (Institute of Physics Publishing, Bristol and 
Philadelphia) p534. See also \verb|http://www1.psi.ch/www_sindrum2_hn/sindrum2.html|.
%
\bibitem{PRISM} Y.~Kuno, presentation at a miniworkshop on Neutrino 
Factories and Muon Storage Rings at CERN, January 17--19 (2000), 
\verb|http://muonstoragerings.web.cern.ch/muonstoragerings/|. 
See also technical notes in the homepage of the PRISM project 
\verb|http://psux1.kek.jp/~prism|.


\end{thebibliography}
\end{document}